%% file: main_arxiv.tex
\documentclass[a4paper,
11pt
]{article}

\usepackage[english]{babel}
\usepackage[a4paper,left=2.5cm,right=2.5cm,top=2.5cm,bottom=2.5cm]{geometry}
\usepackage{fix-cm}
\usepackage[colorlinks=true, urlcolor=blue]{hyperref} 
\usepackage{bm}
\usepackage{authblk}
\usepackage{float}
\usepackage{lineno}
\usepackage{siunitx}
\usepackage[ruled,vlined]{algorithm2e}
\usepackage{algorithmic}
\usepackage{caption}
\usepackage{subcaption}
\usepackage{amssymb} 
\usepackage[dvipsnames]{xcolor}
\usepackage{amsmath, amssymb} 
\usepackage{booktabs}
\usepackage{rotating}
\usepackage{tikz}
\usepackage{colortbl}  
\usepackage{multirow}
\usepackage{array}
\usepackage{url}
\usepackage[square,numbers]{natbib}
\usepackage{lineno}
\usepackage{enumitem}

\graphicspath{ {./images/} }

\title{{Reservoir computing based predictive reduced order model for
steel grade intermixing in an industrial continuous casting tundish}}
\author[1,4]{Harshith~Gowrachari\footnote{hgowrach@sissa.it}}
\author[3]{Mattia~Giuseppe~Barra\footnote{m.barra@danieli.com}}
\author[2]{Giovanni~Stabile\footnote{giovanni.stabile@santannapisa.it}}
\author[3]{Gianluca~Bazzaro\footnote{g.bazzaro@danieli.it}}
\author[1]{Gianluigi~Rozza\footnote{grozza@sissa.it}}

\affil[1]{\small Mathematics Area, mathLab, International School for Advanced Studies\\

Via Bonomea 265, 34136, Trieste, Italy.}

\affil[2]{\small Biorobotics Institute, Sant’Anna School of Advanced Studies\\

V.le R. Piaggio 34, 56025, Pontedera, Pisa, Italy.}

\affil[3]{\small Danieli Research Center, Danieli $\&$ C. S.p.A., Via Nazionale 41, 33042\\

Buttrio, Province of Udine, Italy.}

\affil[4]{Present address: DTIS, ONERA -- The French Aerospace Lab\\

2 Avenue Marc Pélegrin, 31400 Toulouse, France.}

\date{} 

\begin{document}
\maketitle
\begin{center}
\textsuperscript{* \dag} These authors contributed equally.
\end{center}    
    \input{sections/abstract}
    \input{sections/intro}
    \input{sections/methodology}
    \input{sections/domain}
    \input{sections/numerical_method}
    \input{sections/results}
    \input{sections/conclusion}
    \input{sections/ackno}
    \input{sections/appendix}
\bibliographystyle{abbrvnat}
\bibliography{bib/biblio}

\end{document}

%% file: sections/abstract.tex
\begin{abstract}
Continuous casting is a widely adopted process in the steel industry, where maintaining high steel quality is paramount. Efficient prediction of grade intermixing during ladle changeover operations is critical for maintaining steel quality and minimizing material losses in the continuous casting process. Among various factors influencing grade intermixing, operating parameters play a significant role, in addition to tundish geometry and flow control devices. In this study, three-dimensional, transient, two-phase turbulent flow simulations are conducted to investigate the ladle changeover operation. During this process, the molten steel level in the tundish typically varies over time, significantly affecting the grade intermixing phenomena. The influence of ladle change time on intermixing time has been presented. However, high-fidelity full-order simulations of such complex transient phenomena are computationally expensive and are impractical for real-time monitoring or design-space exploration in industrial-scale applications. To address this issue, a reduced order modelling approach based on proper orthogonal decomposition (POD) and reservoir computing (RC) is employed to efficiently predict intermixing time. The proposed reduced order model (ROM) demonstrates excellent predictive accuracy using limited training data while requiring significantly less computational resources and training time. The results demonstrate the potential of the proposed methodology as a fast, reliable tool for real-time process monitoring and optimization in industrial continuous casting operations. \\

\end{abstract}

%% file: sections/intro.tex
\section{Introduction}

\begin{figure}[!ht]
    \centering
    \includegraphics[width=0.75\linewidth]{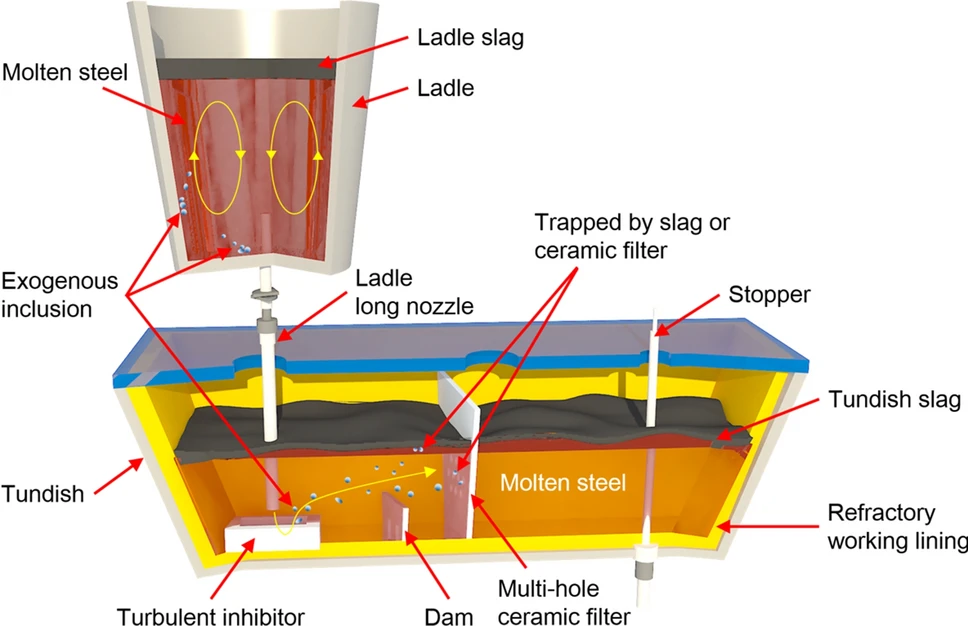}
    \caption[Schematic representation of a single strand tundish]{Schematic representation of single strand tundish \cite{Wang2021}, illustrating the flow physics and key components. The tundish contains molten steel, slag, and inclusion flotation, representing a multiphase flow system. The flow modifiers, such as the impact pod (acting as a turbulent inhibitor) and the dam, improve steel quality and ensure better castability.}
    \label{fig: Tundish}
\end{figure}

Continuous Casting is the predominant method used in steel production, accounting for 96$\%$ of the world's steel production, as per the World Steel Association 2024 report \cite{worldsteel2024}. The first attempts at continuous casting (CC) usage date back to the middle of the 19th century and significantly gained popularity in the 1950s \cite{CCSteelIrvingbook1993}. In the CC machine, the tundish is an intermediate vessel that transfers molten metal evenly from the ladle to the mould, with a desired throughput rate and temperature without inclusion contamination. It acts as a reservoir during the ladle change period and continues to supply molten metal to the mould, when the melt inflow is stopped, making it possible to perform sequential casting based on the availability of the ladles. The tundish plays a significant role in producing clean steel. The single strand tundish in Figure \ref{fig: Tundish} shows the schematic of the flow physics inside the tundish, dealing with multiphase flows.  \\

{{A tundish is a flow reactor}}, made of a refractory-lined channel (rectangular or square cross-section), consisting of one inlet, one or multiple outlets, and sometimes with flow modifiers like an impact pot as a turbulent inhibitor, dams, and ceramic filters placed along the cross-sectional length. It may consist of a refractory-lined lid and outlet ports assembled with stoppers or slide gates to facilitate the melt into the mould. These components of the tundish can be depicted in Figure \ref{fig: Tundish}. The proper design and operation of the tundish play a vital role in producing the desired steel composition and quality, resulting in clean steel production. In the last three decades of the 19th century, significant advancements in tundish technology have occurred for clean steel production \cite{Thomas2017, chattopadhyay2010physical, Mazumdar2018}. However, the advances in the tundish technology have slowed down in recent years, as a result of becoming a mature technology \cite{Mazumdar2018}. It is still of great concern for industries, because of significant interest in developing tundish metallurgy and strict demands regarding steel quality. Also, the concerns of cost, energy, and environmental impacts are becoming significantly important \cite{Siddiqui2019, siddiqui2021mathematical}. \\

In sequential continuous casting of steel, when the teeming ladle is exhausted, it is replaced with a new one, allowing casting to continue without interruption for several hours. However, when the new ladle contains a different steel grade, the mixing of the two grades occurs in the tundish. To prevent grade intermixing, various industrial techniques, such as flying tundish changes and grade separator plates, have been developed for techno-economic reasons.  However, these methods often result in significant revenue losses, increased processing time, and the need for handling multiple tundishes for maintenance and preheating, making them less favourable. With the growing demand for a diverse range of steel products in smaller batches, it has become increasingly important to address these challenges efficiently. The industry has found it more practical to implement a straightforward ladle changeover process, allowing continuous casting of different steel grades using the same tundish. However, this approach leads to the mixing of two distinct steel grades within the tundish, resulting in the formation of an intermediate grade with a composition between the two original grades. \\

Grade intermixing is majorly influenced by tundish operational parameters, such as inflow rate, residual volume, and outflow (throughput) rate have a more pronounced effect on the mixing of grades during sequential casting \cite{Battaglia2012, Siddiqui2014, Siddiqui2018}. These parameters influence the melt flow structure inside the tundish. Understanding and quantifying the mixing process inside the tundish is crucial for steelmakers. It has been observed that the melt bath height in the tundish changes during transient operations \cite{Muralikrishna2013, Siddiqui2014turbulence}. A detailed investigation of the mixing phenomenon inside the tundish was conducted by Odenthal et al.\cite{Odenthal2003}, examining both steady and transient casting conditions, including the transient filling sequence of the tundish and ladle change. They also numerically modelled the tundish filling process using the volume of fluid (VOF) method. Research to date has primarily focused on the steady-state operating bath height of the melt, both through numerical and experimental analysis \cite{Jha2003, Huang1996, Thomas1997, Tripathi2011, Raghavendra2013}. A novel mixing model to quantify the composition distribution in solidified steel during grade mixing is proposed by Cho and Kim \cite{Cho2006}. In their consecutive work \cite{Cho2010}, they demonstrated that the model could predict the intermixed zone for various casting operations, including bloom, slab, and thin slab casting. Alizadeh et al. \cite{Alizadeh2008} developed a numerical model to calculate grade mixing in the tundish and found that the extent of mixed-grade steel depends on the tundish Richardson number. Pieprzyca \cite{pieprzyca2013cast} studied the impact of mixing on the quality of solidified steel, noting that improper flow and mixing methods in the tundish can affect the chemical and thermal homogeneity of the steel, ultimately influencing cast quality. A detailed systematic study of mixing in the tundish was carried out by Muralikrishna et al. \cite{Muralikrishna2013}, who conducted an extensive physical investigation into grade intermixing. They found that residual volume significantly affects mixing and discussed how the internal geometry of the tundish influences grade mixing, developing a correlation to quantify mixed-grade steel formation. Siddiqui and Jha reported \cite{Siddiqui2015} that the inflow rate plays a crucial role in grade mixing during sequential castings. Cwudzinski \cite{Cwudziski2016hydro} conducted a numerical study of flow control devices in the tundish to alter the hydrodynamic patterns, reporting that stagnant volumes could be mitigated by an active recycle zone, and that alloy addition mixing time depends on location and hydrodynamic patterns. Further explored hydrodynamic patterns by developing a numerical model to understand alloy dispersion in the melt, studying residence time distribution (RTD) and alloy dispersion in a single strand tundish. A more comprehensive multiphase study on mixing was conducted by Al-Harbi et al. \cite{AlHarbi2016}. They developed a 3D multiphase numerical model to quantify grade mixing in the tundish, validated by chemical analysis of solidified slabs. They concluded that 80$\%$ of the old steel grade could be removed from the tundish within the first 5-6 minutes of mixed-grade slab casting. Cwudzinski \cite{Cwudziski2016} also conducted physical and numerical experiments to understand the formation of the mixing zone during continuous casting, using the buoyancy number to visualize the hydrodynamic phenomena in the tundish. A recent study by Krashnavtar and Mazumdar \cite{Krashnavtar2018} reported the influence of operating variables on grade intermixing time. More recently, Siddiqui and Kim \cite{Siddiqui2019} conducted numerical simulations to investigate the impact of residual volume and outflow (throughput) rate on the mixing phenomenon inside the tundish. \\

The intermixing of steel grades and the subsequent production of mixed grades occur over a relatively short period, requiring a fully transient formulation for the hydrodynamic modelling of this process. Additionally, a multiphase approach is necessary, as the tundish working volume typically contains two or more phases, which change over time during ladle changeover operations. Mathematical modelling of grade intermixing requires substantial computational resources. Consequently, the studies conducted so far have been largely simplistic and rarely captured the key process features of ladle changeover and intermixing. As a result, despite numerous studies, a physically realistic model of ladle changeover and the associated grade intermixing has not yet been developed \cite{Krashnavtar2018, siddiqui2021mathematical}. \\

Intermixed products are typically downgraded, leading to significant yield and revenue losses \cite{Krashnavtar2018, Siddiqui2019}. As a consequence, the industry has a strong interest in minimizing intermixing within the tundish to reduce the production of mixed-grade slabs or blooms \cite{Krashnavtar2018, Siddiqui2019}. The issue of grade intermixing in steelmaking tundishes has garnered significant attention over the past two decades \cite{Mazumdar1999, chattopadhyay2010physical, Thomas2017, Mazumdar2018, siddiqui2021mathematical}. Efforts have generally focused on predicting intermixing time through both mathematical and physical modelling, both differential and macroscopic models to address this challenge \cite{Krashnavtar2018, siddiqui2021mathematical}. \\

Reduced order models (ROMs) is rapidly gaining prominence in applied mathematics, computational science, and engineering. Industries are increasingly interested in ROMs for engineering systems, particularly in applications related to control, optimization, and uncertainty quantification. These models address the growing demand for efficient computational techniques, especially in many-query scenarios and real-time computations. ROMs have been successfully applied across various fields, including fluid dynamics \cite{rozza2022advanced}, structural dynamics \cite{Jain2017}, and electromagnetic systems \cite{Wittig2006}. They serve as an effective tool for approximating full-order model (FOM) systems, significantly reducing the computational cost required to obtain numerical solutions in parametric settings \cite{quarteroni2015reduced, hesthaven2016certified, rozza2022advanced}. In our recent work, we have explored both projection-based \cite{Gowrachari_PMOR_2509} and data-driven ROM approaches \cite{Gowrachari2026_RE_DDROM} for continuous casting tundish RTD analysis. This work employs a non-intrusive, data-driven ROM to enable efficient and reliable computations. Specifically, the proper orthogonal decomposition with regression (POD-R) \cite{Tezzele2022} strategy is utilized, where POD is used to obtain the reduced basis subspace \cite{hesthaven2016certified, rozza2022advanced}, and artificial neural networks (ANNs) is employed for the regression task to evaluate the modal coefficients \cite{hesthaven2018non, Shah2022, Pichi2023}. \\

Artificial neural networks have become a fundamental and powerful tool in machine learning, particularly well suited for supervised learning in data-driven science and engineering \cite{rozza2022advanced, hartman2017deep, Raissi2019, lee2020model, kim2022fast, bruna2024neural}. They have been widely applied to forecasting problems in complex dynamic systems \cite{Brunton2019, Funahashi1993, hartman2017deep, JinQuanHuang2003, Narendra1990, Lngkvist2014, Li2017}. However, conventional feedforward ANN trained via backpropagation can be computationally expensive to optimize, even with advancements such as stochastic gradient descent and hardware innovations like GPU-based processing. Recurrent neural networks (RNNs) are particularly effective for handling temporal data in dynamical systems \cite{Funahashi1993, Kimura1998, BailerJones1998, Barbounis2006, Rather2015, Choi2016}, as they inherently capture sequential dependencies.  Among these, long short-term memory (LSTM) networks have demonstrated high accuracy and reliability \cite{Chattopadhyay2020, Choi2016, Hochreiter1997, Vlachas2018, Vlachas2020, Yeo2019, Maulik2021, Hajisharifi2024}, but they tend to be data-hungry, training difficulties, requiring long observation times and substantial computational resources for full training \cite{Razvan2013, Kutz2017, Gauthier2021}. \\

The reservoir computing (RC) \cite{Jaeger2004, Verstraeten2007, Lukoeviius2009}, along with its closely related models, the echo state network (ESN) \cite{Jaeger2001, Lukoeviius2012} and liquid state machine (LSM) \cite{Maass2002, Grzyb2009}, represent the specialized variants of RNNs in which only the output layer is trained, while the internal network weights remain fixed. This approach significantly reduces the computational complexity, as it requires only a simple and efficient least-square computation rather than the costly non-linear optimization typically needed for a fully trained RNN. A particularly striking outcome is that, despite this substantial simplification, RC can still achieve competitive forecasting performance, even for chaotic or spatio-temporally complex systems \cite{Canaday2018, Carroll2019, Daniel_ReservoirComputing_SIAMNews, Lu2018, Pathak2017, Pathak2018, Zimmermann2018}. RC is especially effective when a full-state observation is available, whereas more sophisticated RNN variants, such as LSTMs, generally outperform it when only a reduced set of variables is accessible \cite{Vlachas2018, Vlachas2020}. Nonetheless, RC remains widely adopted, not only due to its computational efficiency and ease of training but also because of its surprising accuracy \cite{Bollt2021}. \\

Full-order simulations of steel grade intermixing during ladle changeover operations, which involve multiphase flows, are computationally expensive and require long-time integration, as these operations typically last for 10-15 minutes (physical time). This high computational cost hinders parametric studies and limits the availability of training data for surrogate modelling. To overcome these challenges, we develop a POD and RC based reduced order modelling framework. \\

In this study, the POD-RC-ROM approach is used to predict the intermixing time during ladle changeover operations. The framework integrates the POD and RC model, where POD is utilized to obtain a low-dimensional representation of the tundish system dynamics, and RC is employed as an efficient non-linear regression technique to approximate the temporal evolution of modal coefficients. Integrating reservoir computing architectures into ROM frameworks presents multiple advantages. A primary benefit lies in the significantly lower training cost, as only the output weights are trained, eliminating the need for backpropagation. RC is also well-suited for capturing temporal dynamics, leveraging the reservoir to project inputs into a high-dimensional space through the reservoir. Among RC approaches, ESNs are particularly notable for their echo state property, which ensures long-term stability and efficiency in modelling complex dynamical systems. Compared to conventional deep learning models such as Recurrent Neural Networks (RNNs), Long Short-Term Memory (LSTM) networks \cite{Razvan2013} and Deep Neural Networks (DNNs) \cite{Kutz2017}, RC methods offer comparable accuracy while requiring significantly fewer training data and computational resources \cite{Gauthier2021}. As reported in \cite{Bollt2021}, RC models can achieve remarkable predictive performance with reduced training requirements. By leveraging the computational efficiency and accuracy of the POD-RC-ROM framework, this study aims to enable fast and reliable predictions for real-time monitoring and optimization of ladle changeover operations. \\

The main contributions of this work are as follows:

\begin{itemize}
    \item We develop a non-intrusive reduced order modelling framework based on proper orthogonal decomposition and reservoir computing (i.e., POD-RC-ROM) to predict grade intermixing during ladle changeover in continuous casting processes. Specifically, we leverage Echo State Networks (ESNs) to capture the temporal dynamics of the modal coefficients, enabling efficient training without backpropagation. This approach simplifies the training process while maintaining high predictive accuracy, as demonstrated through comparisons with traditional machine learning architectures \cite{Bollt2021}.

    \item We demonstrate that the proposed POD-RC-ROM achieves accurate predictions with limited training data, significantly reducing both the computational cost and training time. The framework's efficiency and scalability make it well-suited for real-time industrial applications involving complex multiphysics phenomena.

    \item We evaluate the POD-RC-ROM in the extrapolation regime by comparing its predictions against high-fidelity, three-dimensional, transient, multiphase flow simulations under industrially relevant conditions. The reduced order model accurately captures the grade intermixing occurring during ladle changeover operations, demonstrating its robustness and generalization capability beyond the training regime even with limited training data. 

    \item We present a computationally efficient and scalable modelling framework that enables real-time digital twins and process monitoring, optimization, and design-space exploration in continuous casting. This offers a practical tool to support more sustainable and intelligent steelmaking operations.

\end{itemize}

This paper is organised as follows. In Section~\ref{sec:Physicalproblem}, we explain the physical problem under consideration; in Section~\ref{sec:FOM}, we detail the full-order model, including the governing equations, the VOF method, turbulence modelling, and species transport; in Section~\ref{sec:ROM}, we present the reduced order modelling framework based on proper orthogonal decomposition and reservoir computing, including the echo state network and the evaluation of modal coefficients; in Section~\ref{sec:num_setup}, we describe the numerical setup, comprising the computational domain and boundary conditions; in Section~\ref{sec:num_method}, we outline the numerical methods employed; in Section~\ref{sec:result}, we present and discuss the results obtained from both the full and reduced order models; and finally, conclusions and perspectives are discussed in Section~\ref{sec:conslusion}.

\section{Physical problem}\label{sec:Physicalproblem}
In this work, we are particularly focusing on transient operations occurring during the ladle over. Once the molten steel or the melt in the ladle is exhausted, it is replaced with a new ladle to maintain sequential casting. During this transition, the inflow ceases for a period, and as a result, the melt level drops as there is a constant throughput rate required for continuous casting. As the new ladle delivers melt into a tundish, it can lead to severe thermal and material mixing between melt, slag, and the ambient air, which can lead to air entrainment and slag emulsification. 
\begin{figure}[ht!]
    \centering
    \includegraphics[width=0.85\linewidth]{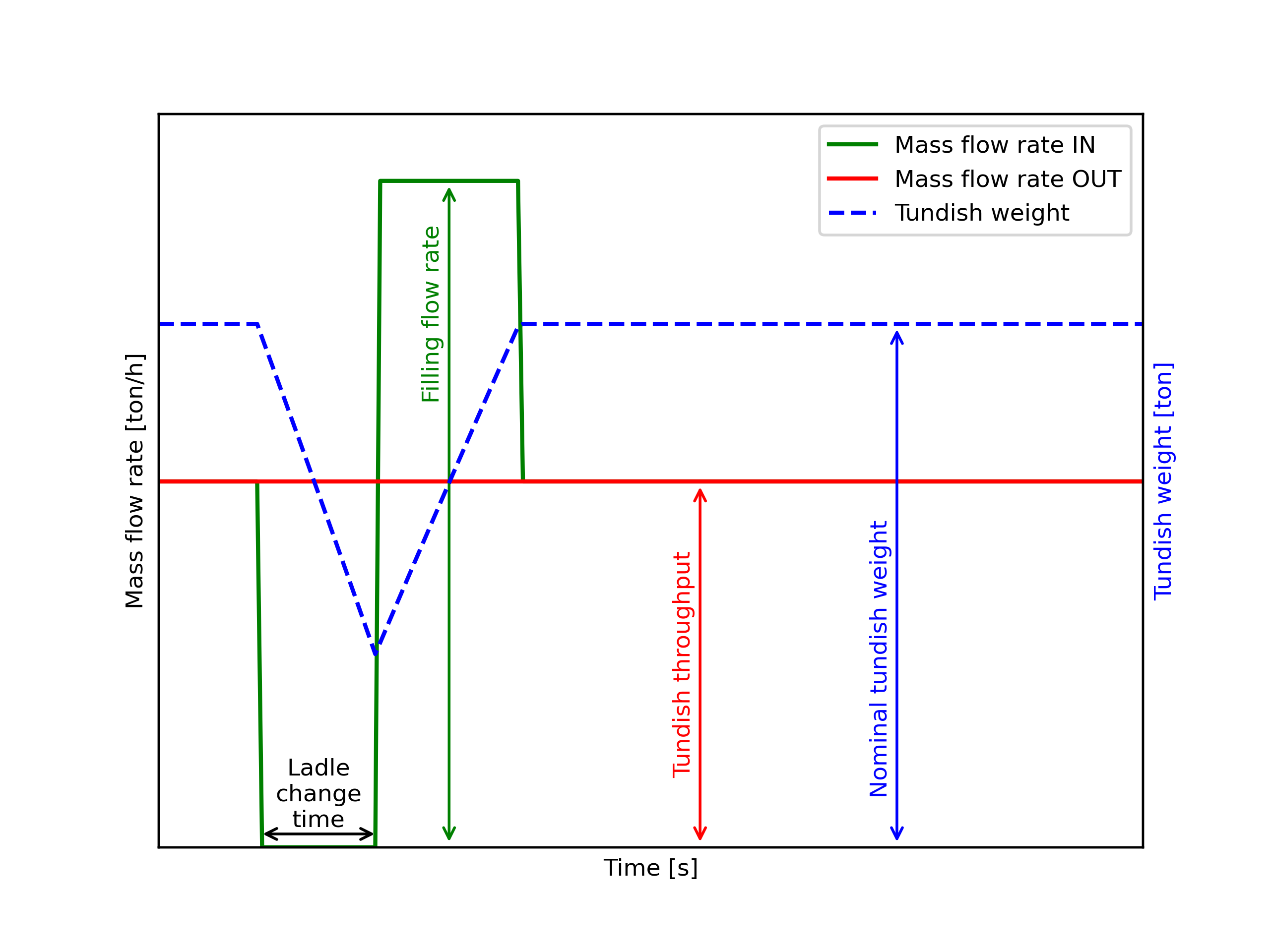}
    \caption{Process diagram of the ladle changeover operation}
    \label{fig:ladleChangeOverOperations}
\end{figure}

Figure \ref{fig:ladleChangeOverOperations} illustrates the different stages of the ladle change operation. Throughout this operation, the volume inside the tundish changes continuously. At any instant of time, the working volume of a tundish is occupied by two or more phases in variable proportions, changing with time. The ladle changeover and the resulting grade intermixing can be categorized into three distinct yet interconnected stages of tundish operation:

\begin{enumerate}[leftmargin=*, label=\textbf{Stage \arabic*}:]
    \item Steady-state operation before ladle changeover.
    \item Transient emptying phase with outflow but no inflow.
    \item Transient refilling phase with a new ladle containing a new steel grade.
\end{enumerate}

The duration of Stage 1 varies based on the ladle size and casting rate and can extend for an hour or longer. In contrast, Stages 2 and 3 together typically last no more than 10–15 minutes before the process stabilizes and returns to steady-state operation. \\

Tundish metallurgical operations primarily involve multiphase flow, heat and mass transfer, turbulence, and chemical interactions between various phases, including molten metal, slag, refractory material, and ambient air. Accurately modelling these processes is complex and presents several challenges. \\

In this study, we focus on modelling a single-strand tundish. For the numerical investigation, we simulate three-dimensional, transient, two-phase, isothermal, and turbulent flow dynamics. The key quantity of interest (QoI) in this process is steel grade intermixing, particularly the new steel grade transition at the outlet of the tundish. To simulate these phenomena sequentially, the following assumptions have been made:

\begin{itemize}

    \item The tundish is modelled as a two-phase flow system, considering molten steel and slag as the two immiscible fluids with fixed physical properties, while the presence of air is neglected.
    
    \item The fluid flow is assumed to be isothermal.  

    \item Flow phenomena within the ladle shroud are assumed to be homogeneous.  
    
    \item There are no chemical interactions between the phases within the molten bath.
    
    \item Only half of the 0.5-scaled industrial tundish was modelled, taking advantage of its geometric symmetry.
    
\end{itemize}

{{Further, we highlight that the region above the liquid steel is entirely filled with slag. This simplification avoids the complexity associated with three-phase flow modelling, as the primary focus of this work is on grade intermixing in the steel phase.}}

%% file: sections/methodology.tex
\section{Full order model}\label{sec:FOM}

Multiphase modelling in tundish metallurgical operations enables the investigation of key phenomena such as molten steel–slag interface dynamics during ladle changeover processes. In the context of sequential casting, understanding and quantifying steel grade intermixing is particularly important for quality control and process optimisation. This section presents the full-order model considered to resolve the coupled flow, interface evolution, turbulence, and species transport. The VOF method is employed to capture the transient melt surface development during ladle changeover, which directly influences the extent of grade intermixing.

\subsection{Governing equations}
To model steel grade intermixing in the tundish under isothermal conditions, 
Unsteady Reynolds-averaged Navier–Stokes (URANS) equations are employed, which are given by
\begin{equation}
\left\{
\begin{aligned}
    &\frac{\partial \rho}{\partial t } + \nabla \cdot (\rho \mathbf{U}) = 0, 
    && \text{in } \Omega \times (0,T], \\[6pt]
    &\frac{\partial (\rho \mathbf{U})}{\partial t } 
    + \nabla \cdot (\rho \mathbf{U} \otimes \mathbf{U}) 
    = -\nabla p 
    + \nabla \cdot \left[ \mu_{\text{eff}} (\nabla \mathbf{U} + \nabla^{T} \mathbf{U}) \right] 
    + \mathbf{F}, 
    && \text{in } \Omega \times (0,T], \\[6pt]
    & \text{boundary condition}, 
    && \text{on } \Gamma \times (0,T], \\[6pt]
    & \text{initial condition}, 
    && \text{in } \Omega, \\[6pt]
\end{aligned}
\right.
\label{eqn: RANS}
\end{equation}
where $\rho$ is the density, $\mathbf{U}$ is the velocity, $\textit{p}$ is the pressure and $\mathbf{F}$ denotes gravitational and external body forces. 
The problem is defined in a spatial domain $\Omega \subset \mathbb{R}^3$ over the time interval $t \in [0, T]$, and is subjected to boundary conditions on $\Gamma$ and prescribed initial conditions.

\subsection{Volume of Fluid method}

To capture the interface between two immiscible fluids, the Volume of Fluid (VOF) method based on the one-fluid model is employed. 
In this formulation, both fluids share a single velocity field, i.e., $\mathbf{U} = \mathbf{U}_{1} = \mathbf{U}_{2}$. 
The physical properties of the fluids are averaged over the domain, which leads to constituent relations of $\rho$ and $\mu$ weighted values given by:  

\begin{equation}
\begin{aligned}
    \rho &= \alpha_1 \rho_1 + (1-\alpha_1)\rho_2, \\
    \mu  &= \alpha_1 \mu_1  + (1-\alpha_1)\mu_2,
\end{aligned}
\end{equation}
where $\rho_1, \rho_2$ and $\mu_1, \mu_2$ denote the constant densities and dynamic viscosities of the two immiscible fluids. \\

In the VOF method, the modified volume fraction advection equation (\ref{eqn: adv eqn}) is solved simultaneously with continuity and momentum equations. Where $\alpha \in [0,1]$ is phase fraction which takes values $\alpha = 0$ for volume is filled only with phase 2, $\alpha = 1$ for control volume is filled only with phase 1 and at the interface, it ranges between $ 0 < \alpha < 1 $, and $\mathbf{U}_r$ is the relative velocity shown in (\ref{eqn: relative velocity}) and is perpendicular to the interface.
 
\begin{equation}
\frac{\partial \alpha}{\partial t} + \nabla \cdot\left(\alpha \mathbf{U} \right) + \nabla \cdot\left[\mathbf{U}_r \alpha\left(1-\alpha\right)\right] = 0, 
\label{eqn: adv eqn}
\end{equation}

\begin{equation}
\mathbf{U}_r = C_\alpha|\mathbf{U}| \frac{\nabla \alpha}{\left|\nabla \alpha \right|}, \hspace{8mm} \text{with} \hspace{2mm}\mathcal{C}_\alpha \in [0,1].
\label{eqn: relative velocity}
\end{equation}

\subsection{Turbulence model}
For turbulence modelling in the tundish, we consider the $k-\varepsilon$ turbulence model based on \cite{launder1983numerical}, which is a two-equation linear eddy viscosity turbulence closure model consisting of $k$ - turbulent energy and $\varepsilon$ - turbulent kinetic energy dissipation rate equations: 

\begin{equation}
\frac{\partial (\rho k)}{\partial t} = \nabla \cdot \left( \rho D_k \nabla k \right) + P - \rho \varepsilon, 
\label{Eqn:k-eqn}
\end{equation}

\begin{equation}
\frac{\partial (\rho \varepsilon)}{\partial t} = \nabla \cdot\left(\rho D_\varepsilon \nabla \varepsilon\right) + \frac{C_1 \varepsilon}{k} \left(P + C_3 \frac{2}{3} k \nabla \cdot \mathbf{U} \right) - C_2 \rho \frac{\varepsilon^2}{k}, 
\label{Eqn:epsilon-eqn}
\end{equation}

\begin{equation}
    \mu_t = \rho C_\mu \frac{k^2}{\varepsilon},  
    \label{Eqn:turbulent viscosity}
\end{equation}

\noindent where, $D_k$ - effective diffusivity for $k$, $P$ - turbulent kinetic energy production rate, $D_{\varepsilon}$ - effective diffusivity for $\varepsilon$, $C_1, C_2, C_3$ are model coefficients. After obtaining $k$, $\varepsilon$,  values, the turbulent (eddy) viscosity, $\mu_{t}$ is given by (\ref{Eqn:turbulent viscosity}), here $C_\mu$ is the model coefficient for turbulent viscosity. We consider the standard model coefficient values shown in Table \ref{tab: model coefficents}. 

\begin{table}[h]
    \centering
    \begin{tabular}{ccccc}
        $C_\mu$ & $C_1$  & $C_2$  & $C_3$  & $\sigma$ \\
        \hline \hline
        0.09 & 1.44  & 1.92 & 0 & 1  \\
        \hline 
    \end{tabular}
    \caption{Model coefficients of $k-\varepsilon$ turbulence model.}
    \label{tab: model coefficents}
\end{table}

\subsection{Species transport equation}
The evolution of the new steel grade is modelled using the species transport equation (\ref{eqn: species transport equation}), which is a convection-diffusion equation:

\begin{equation}
    \frac{\partial(\rho C)}{\partial t} + \nabla \cdot (\rho \mathbf{U} C) =
    \nabla \cdot (\mu_{eff} \nabla C), 
    \label{eqn: species transport equation}
\end{equation}
\begin{equation}
    \mu_{eff} = \mu + \mu_t,  
    \label{eqn:effective_diffusivity}
\end{equation}
where, \textit{C} denotes the concentration of new steel grade, $\mu_{eff}$ represents the effective diffusivity (\ref{eqn:effective_diffusivity}), defined as the sum of molecular diffusivity - $\mu$ and turbulent diffusivity - $\mu_t$.

\section{Reduced order model}\label{sec:ROM}

This section presents the reduced order modelling framework developed to efficiently approximate the dynamics of the full-order model. The approach combines POD with RC, resulting in a data-driven method capable of capturing the essential flow dynamics at a significantly reduced computational cost. In~\ref{subsec:POD}, the POD method is introduced to extract dominant spatial modes from high-fidelity simulation data. In~\ref{subsec:RC}, the RC framework is discussed, with emphasis on the ESN and the procedure for evaluating modal coefficients. Finally, in~\ref{subsec:Algorithm}, the overall POD-RC-ROM algorithm is described, outlining both the offline training and the online prediction stage. \\

In this work, we integrate POD with RC to develop a data-driven ROM \cite{Tezzele2022}. The framework consists of two stages:

\begin{itemize}
\item \textbf{Offline stage}: First, full-order high-fidelity solutions are obtained by solving the FOM for parameters of interest. Following this, a reduced basis subspace is obtained by performing POD on a set of high-fidelity snapshots. Once the reduced basis subspace is obtained, the original snapshots are projected onto this reduced subspace to get the corresponding parameter-dependent modal coefficients. Subsequently, reservoir computing is then employed to approximate the mapping between parameters and modal coefficients. This stage is computationally expensive but is performed only once. 
    
\item \textbf{Online stage}: Given a new parameter, the trained reservoir computing model evaluates the corresponding modal coefficients. The reduced order solution is then reconstructed as a linear combination of the POD basis functions weighted by these modal coefficients. This enables efficient exploration of the parameter space at a significantly lower computational cost.
\end{itemize}

In the context of a parameter-time dependent problem, the collection of all possible full-order model (FOM) solutions can be described by the solution manifold $\mathcal{M}$. This manifold is defined over the temporal domain $\mathcal{T}$ and parameter space $\mathcal{P}$ as:
\begin{equation}
\mathcal{M} := \left\{\, C(t; \boldsymbol{\mu}) \;\middle|\; (t, \boldsymbol{\mu}) \in \mathcal{T} \times \mathcal{P} \,\right\}.
\end{equation}

Its discrete counterpart, denoted $\mathcal{M}_h$, is defined as the collection of high-fidelity numerical solutions:
\begin{equation}
\mathcal{M}_h := \left\{\, C_{N_h}(t; \boldsymbol{\mu}) \;\middle|\; (t, \boldsymbol{\mu}) \in \mathcal{T} \times \mathcal{P} \,\right\} \subset \mathbb{R}^{N_h},
\end{equation}
where $N_h$ denotes the number of degrees of freedom. In other words, $\mathcal{M}_h$ represents the high-dimensional discrete solution space explored by the FOM across all time and parameter variations. \\

Let us consider a set of parameters values \( \mathcal{K} = \{\boldsymbol{\mu}_1, \dots, \boldsymbol{\mu}_{N_p}\} \subset \mathcal{P} \), a finite training set sampled from the parameter domain \( \mathcal{P} \). For each parameter \( \boldsymbol{\mu}_k \in \mathcal{K} \), we denote a set of time instances \( t_j^{(k)} \in \{t_1^{(k)}, \dots, t_{N_t^{(k)}}^{(k)} \}  \in (0, T] \subset \mathcal{T} \), where the number of time instances \( N_t^{(k)} \) may vary for each parameter, drawn from the time domain $\mathcal{T}$. The FOM is evaluated for each \( \boldsymbol{\mu}_k \), producing a different number of snapshots per parameter. Consequently, the total number of snapshots is given by \( N_s = \sum_{k=1}^{N_p} N_t^{(k)} \). Once the full-order solutions are computed, they are assembled column-wise in a snapshots matrix \( \mathcal{S} \in \mathbb{R}^{N_h \times N_s} \), as follows:

\begin{equation}
    \mathcal{S} = 
    \left[\begin{array}{cccccc}
    C_1\left(t_1; \boldsymbol{\mu}_1\right) & \cdots & C_1\left(t_{N_t^{(1)}}; \boldsymbol{\mu}_1\right) & C_1\left(t_1; \boldsymbol{\mu}_2\right) & \cdots & C_1\left(t_{N_t^{(N_k)}}; \boldsymbol{\mu}_{N_k}\right) \\
    \vdots & \vdots & \vdots & \vdots & \ddots &  \vdots \\
    C_{N_h}\left(t_1; \boldsymbol{\mu}_1\right) & \cdots & C_{N_h}\left(t_{N_t^{(1)}}; \boldsymbol{\mu}_1\right) & C_{N_h}\left(t_1; \boldsymbol{\mu}_2\right) & \cdots & C_{N_h}\left(t_{N_t^{(N_k)}}; \boldsymbol{\mu}_{N_k}\right)
\end{array}\right]
\end{equation}

\subsection{Proper orthogonal decomposition}\label{subsec:POD}

POD is a widely used linear approximation method in reduced order modelling, particularly in computational fluid dynamics applications over the past decades \cite{rozza2022advanced}. In this work, POD is employed to derive an optimal orthonormal basis and construct a reduced basis subspace that effectively approximates the full-order solutions. The orthonormal basis is obtained by computing POD modes using singular value decomposition (SVD) \cite{PODstewart1993early}, as follows:
\begin{equation}
    \underbrace{\mathcal{S}}_{N_h \times N_s} = 
    \underbrace{\mathbf{U}}_{N_h \times N_h} 
    \underbrace{\boldsymbol{\Sigma}}_{N_h \times N_s} 
    \underbrace{\mathbf{V}^T}_{N_s \times N_s} 
    \Rightarrow 
    \underbrace{\mathcal{S}}_{N_h \times N_T} \approx 
    \underbrace{\hat{\mathbf{U}}}_{N_h \times r} 
    \underbrace{\hat{\boldsymbol{\Sigma}}}_{r \times N_s} 
    \underbrace{\hat{\mathbf{V}}^T}_{N_s \times N_s}, 
    \label{eqn:svd} 
\end{equation}
\noindent where, $\mathbf{U} = \left\{ \varphi_1|\dots|\varphi_{N_h} \right\} \in \mathbb{R}^{{N_h} \times {N_h}}$ and $\mathbf{V} = \left\{ \mathbf{w}_1|\dots|\mathbf{w}_{N_s} \right\} \in \mathbb{R}^{{N_s} \times {N_s}} $ are orthonormal matrices. The Columns of $\mathbf{U}$ are the left singular vectors, also known as POD modes, and the columns of $\mathbf{V}$ are the right singular vectors of the snapshot matrix $\mathcal{S}$. The diagonal matrix $\mathbf{\Sigma} \in \mathbb{R}^{N_h \times N_s}$ contains \textit{L} non-zero real singular values arranged in descending order $\sigma_{1} \geq \sigma_{2} \geq \cdots \geq \sigma_{L}>0$, indicating the energy contribution of the corresponding modes. According to the Schmidt–Eckart–Young–Mirsky theorem \cite{eckart1936approximation}, the objective is to approximate the column of $\mathcal{S}$ using a reduced basis subspace spanned by the first $r<L$ dominant left singular vectors, thus identifying a low-dimensional representation of the original data.

\begin{equation}
    \mathcal{V} = \text{span}\left\{ \varphi_i \right\}_{i=1}^{r} \subset L^{2}(\Omega). 
\end{equation}
The reduced basis subspace is obtained by selecting the first \textit{r} singular values and their corresponding left singular vectors, also referred to as POD modes. The resulting POD modes satisfy the following minimization problem: 
\begin{equation}
    \operatorname*{argmin}_{\mathcal{V} \in \mathbb{R}^{N_h \times r}} \left\|\mathcal{S} - \mathcal{V} \mathcal{V}^T \mathcal{S}\right\|_F, \quad \text{where} \quad \mathcal{V}^T \mathcal{V} = \mathbb{I},
\end{equation}
where $ \left\|  \cdot \right\|_F $ represents the Frobenius norm, $\mathcal{V}$ is the reduced basis subspace of dimension \textit{r} and $\mathcal{S}$ is the snapshots matrix. Thus, the reduced basis subspace is the set of vectors that minimize the distance between the original snapshots and their projection onto the subspace spanned by the POD modes. \\

The energy retained by the first \textit{r} POD modes is determined by the expression in (\ref{eqn: POD tolerance}), where r is chosen such that the cumulative energy \( E(r) \)  exceeds a specified threshold.
\begin{equation}
    E(r) = \frac{\sum_{i=1}^r \sigma_i^2}{\sum_{i=1}^{L} \sigma_i^2}.
    \label{eqn: POD tolerance}
\end{equation}
Once the reduced basis subspace has been constructed, the reduced order solution \( C_{rb}(t_k, \boldsymbol{\mu}_k) \) is obtained, which serves as an approximation to the high fidelity full order solution \( C_h(t_k, \boldsymbol{\mu}_k) \):
\begin{equation}
    C_h(t_k; \boldsymbol{\mu}_k) \approx C_{rb}(t_k; \boldsymbol{\mu}_k) = \sum_{j=1}^r [\mathcal{V}^T C_h(t_k;\boldsymbol{\mu}_k)]{_j} \varphi_j, 
\end{equation}
where $(\mathcal{V}^T C(t_k, \boldsymbol{\mu}_k)){_j}$ is the modal coefficient corresponding to the \textit{j}-th mode. 

\subsection{Reservoir computing}\label{subsec:RC}

Reservoir computing is a machine learning framework designed to efficiently capture temporal dependencies in dynamical systems.  It focuses primarily on input-output relationships and is a branch of neuromorphic computing, which mimics the brain's efficient information processing at low energy costs \cite{Jaeger2004, Tanaka2019}. From a machine learning perspective, RC models fall under the category of RNNs and are particularly effective for modelling dynamical systems. RC excels in tackling even the most difficult problems, including those involving chaotic \cite{Pathak2017} or complex spatio-temporal behaviours \cite{Pathak2018}, which an optimized RC can process with ease. \\

The RC technique has been developed through three distinct methods: liquid state machine, echo state network, and backpropagation decorrelation learning rule. Unlike traditional recurrent neural networks, which often require high computational cost and complex training procedures when applied to non-linear dynamical systems. RC offers an effective alternative. It leverages a fixed, randomly initialized dynamical system as a reservoir, with the training occurring only on the output layer via a regularized linear least-squares optimization procedure. The typical RC consists of three components: 
\begin{itemize}
    \item Input layer: Maps input data into the reservoir, often applying a simple transformation or projection. 

    \item Reservoir layer: A high-dimensional, fixed, recurrent dynamical system that transforms the input data into complex, temporal patterns.

    \item Output layer: A trainable layer that maps the updated reservoir state to the desired output, typically using linear regression or other techniques.

\end{itemize}

\begin{figure}[ht!]
    \centering
    \includegraphics[width=0.5\linewidth]{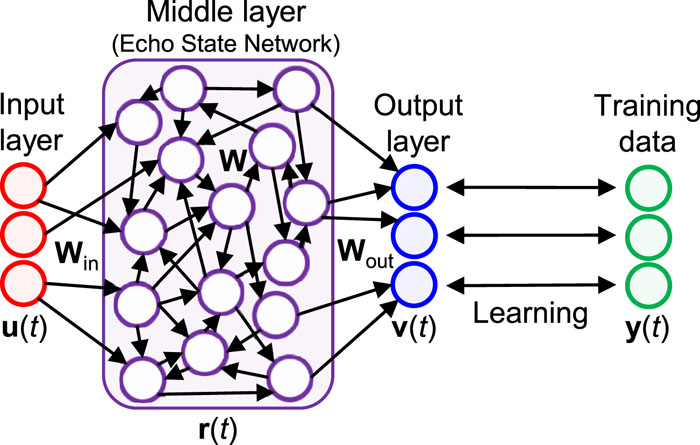}
    \caption{Schematic representation of an echo state network, a type of reservoir computing (RC), as defined in (\ref{eqn:reservoir_update_rule}). The network consists of a randomly initialized \( N_\text{r} \times N_{\text{in}} \) input weight matrix \( W_{\text{in}} \) that maps the \( N_\text{in} \times 1 \) input state vector \( \mathbf{u}(t) \) into the reservoir. The reservoir dynamics are governed by a randomly selected \( N_\text{r} \times N_\text{r} \) recurrent weight matrix \( \mathbf{W} \), which updates the internal reservoir state vector \( \mathbf{r} \) of size \( N_\text{r} \times 1 \). Finally, the trained readout weight matrix \( \mathbf{W}_{\text{out}} \), of dimensions \( N_\text{out} \times N_\text{r} \), is used to produce the final output.}
    \label{fig:RC schematics}
\end{figure}

\subsubsection{Echo state network}\label{subsubsec:RC}
In this work, we consider the echo state recurrent neural networks, the schematic is presented in Figure \ref{fig:RC schematics} \cite{Iwasaki2023}. An echo state network (ESN) is a random recurrent neural network. Given a predefined input \( \mathbf{u}(t) \in \mathbb{R}^{N_{\text{in}}} \) and target output \( \mathbf{y}(t) \in \mathbb{R}^{N_{\text{out}}} \), the node state \( \mathbf{r}(t+1) \in \mathbb{R}^{N_\text{r}} \) and the corresponding output \( \mathbf{v}(t+1) \in \mathbb{R}^{N_{\text{out}}} \) are computed as follows:

\begin{equation}
\mathbf{r}(t + 1) = \underbrace{(1-\gamma)\mathbf{r}(t)}_{\text{Linear Memory}} + \underbrace{\gamma f \left[\beta\left(\varepsilon \mathbf{W}_{\text{in }} \mathbf{u}(t)+\mathbf{W} \mathbf{r}(t)+\eta \mathbf{b}\right)\right]}_{\text{Nonlinear Activation}}, 
\label{eqn:reservoir_update_rule}
\end{equation}
\begin{equation}
    \mathbf{v}(t+1)=\mathbf{W}_{\text{out}} \mathbf{r}(t+1), 
    \label{eqn:reservoir_output_rule}
\end{equation}
where $f: \mathbb{R} \mapsto \mathbb{R}$ is the non-linear activation function, drawing from terminology in the machine learning and neural network literature to replicate the behaviour of a biological network that activates when a certain voltage threshold is reached. The most common choice of \textit{f} would be tanh$(\cdot)$, which applies the scalar hyperbolic tangent function element-wise. Other activation functions, such as sigmoid and the ReLU functions, are widely used in neural network theory, although the ReLU is less commonly employed in RC \cite{Buehner2006}. \( \mathbf{b} \in \mathbb{R}^{N} \) is a constant bias vector, initialized as a random vector following a standard normal distribution. The parameters \( \beta \in \mathbb{R} \), \( \varepsilon \in \mathbb{R} \), and \( \eta \in \mathbb{R} \) are the hyperparameters governing the dynamics of the RC model. Specifically, $\beta$ is a scaling factor that adjusts the amplitude of the input to the activation function, $\varepsilon$ is a scaling factor applied to the input-to-reservoir weights, controlling the influence of the input signal on the reservoir state, and $\eta$ scales the contribution of the bias vector $\mathbf{b}$ to the reservoir dynamics.  The dimensions \( N_{\text{in}} \), \( N_\text{r} \), and \( N_{\text{out}} \) denote the number of input features, reservoir node states, and output features, respectively. The reservoir maps input \( u(t) \) into \( N_\text{r} \) dimensional space, where a sufficiently large \( N_\text{r} \) ensures accurate computation of \( \mathbf{v}(t) \) by ESN. The weight matrices connecting the input to the reservoir, \( \mathbf{W}_{\text{in}} \in \mathbb{R}^{N_\text{r} \times N_{\text{in}}} \), and those within the reservoir, \( \mathbf{W} \in \mathbb{R}^{N_\text{r} \times N_\text{r}} \), are initialized randomly, typically following a standard normal distribution and remain fixed during training and testing phase. As highlighted in (\ref{eqn:reservoir_update_rule}), the right-hand side consists of two components: a linear memory term and a nonlinear activation term, which are combined via a leakage or decay rate \( 0 \leq \gamma \leq 1\). $\mathbf{W}$ should be designed with specific properties, such as spectral radius for convergence \cite{Carroll2019, Jiang2019}, sparsity \cite{Lu2018, Pathak2018, Pantelis2019}, or other considerations related to the echo state property \cite{Buehner2006}. In RC, the only trainable parameter is the weight matrix \( \mathbf{W}_{\text{out}} \in \mathbb{R}^{N_{\text{out}} \times N_\text{r}} \) between the reservoir and the output layer. \\

The output layer represent the RC output \( \mathbf{v}(t+1) = y (t+1) \) as a linear transformation of \( \mathbf{W}_{\text{out}}\) and the reservoir state $\mathbf{r}(t+1)$, as shown in (\ref{eqn:reservoir_output_rule}). The RC model is trained using supervised learning through regularized least-squares regression. The optimal \( \mathbf{W}_{\text{out}} \) is determined by solving the following optimization problem:
\begin{equation}
    \underset{\mathbf{W}_{\text {out }}}{\arg \min }  \sum_{t=1}^n\left\|\mathbf{y}(t)-\mathbf{W}_{\text {out }} \mathbf{r}(t)\right\|_2^2+ {\lambda}\left\|\mathbf{W}_{\text {out }}\right\|_F^2,
    \label{eqn:RC_optimization} 
\end{equation}
where \( \lambda \) is a regularization parameter that prevents overfitting to the training data. The solution to (\ref{eqn:RC_optimization}) is obtained in the least squares sense using Tikhonov regularization, as follows:  

\begin{equation}
    \mathbf{W}_{\text {out }}=\sum_{t=1}^n \mathbf{y}(t) \mathbf{r}(t)^{\top}\left(\lambda \mathbf{I}+\sum_{t=1}^n \mathbf{r}(t) \mathbf{r}(t)^{\top}\right)^{-1}
\end{equation}

This approach significantly reduces the computational cost of training compared to conventional neural network architectures, as only the output weights are optimized by a simple and efficient least squares computation.  

\subsection{Evaluation of modal coefficients}\label{subsec:modal_coeffecients}

The modal coefficients are obtained via projection $\mathcal{V}^T C_{N_h}\left(t_k; \boldsymbol{\mathbf{\mu}}_k \right)$. As mentioned earlier, we employ the RC model to obtain a reliable approximation, $\mathbf{\Theta}$, for the following input-output relation: 
\begin{equation}
\mathbf{\Theta} : \left(t_k; \boldsymbol{\mathbf{\mu}}_k\right) \in \mathbb{R}^{N_s} \mapsto \left[\mathcal{V}^T C_{N_h}\left(t_k; \boldsymbol{\mathbf{\mu}}_k \right) \right]_{j=1}^r \in \mathbb{R}^r.
\end{equation}

During the online phase, the reduced solution for any new time instant $ t_{new} $ and new parameter $ \boldsymbol{\mu}_{new} $ can be computed efficiently as follows:
\begin{equation}
C_{r b}\left(t^*; \boldsymbol{\mu}^{*}\right)=\sum_{j=1}^r \mathbf{\Theta}_j\left(t^{*}; \boldsymbol{\mu}^{*} \right) \mathbf{\varphi}_j, 
\end{equation}
here, \( \mathbf{\Theta}_j\left(t^*; \boldsymbol{\mu}^{*} \right) \) represents the modal coefficients at the new time instance and parameter, with each \( \mathbf{\varphi}_j \) corresponding to the \( j \)-th basis function or mode. 

\subsection{POD-RC-ROM algorithm}\label{subsec:Algorithm}
Algorithm \ref{Algorithm1} outlines the complete framework for the construction of POD-RC based ROMs. The offline stage involves computing full-order solutions for parameters of interest and the construction of ROM, including the training of the RC model to build a non-linear regression of the map $\mathbf{\Theta}$. By leveraging the inherent separation between the training and evaluation phases of RC models, we effectively satisfy the prerequisite for ensuring the decoupling of offline-online stages and computational efficiency during the online stage.  Once trained, the online stage rapidly predicts reduced order solutions for new input parameters with minimal computational cost. This framework enables accurate and efficient approximations, making real-time inference of the QoI feasible. \\

\begin{algorithm}

\caption{POD-Reservoir Computing based ROM}
\label{Algorithm1}
\begin{algorithmic}[1]
\vspace{2mm}
    \STATE \textbf{Offline Stage:}
    \STATE Collect the parameter set \( \mathcal{K} = \{\boldsymbol{\mu}_1, \dots, \boldsymbol{\mu}_{N_p}\} \);
    \STATE Compute full-order solutions:  
    \(
    \{C_{N_h}(t_j^{(1)}; \boldsymbol{\mu}_{1}), \dots, C_{N_h}(t_j^{(N_p)}; \boldsymbol{\mu}_{N_p})\};
    \)
    \vspace{1mm}
    \STATE Extract POD basis functions \( \{\varphi_1,\dots,\varphi_r\} \) via Singular Value Decomposition (SVD);
    \vspace{1mm}
    \STATE Construct the reduced basis subspace:  
    \(
    \mathcal{V} = \left\{ \varphi_1,\dots,\varphi_r \right\};
    \)
    \vspace{1mm}
    \STATE Perform projection to evaluate modal coefficients;
    \vspace{1mm}
    \STATE Train the reservoir computing (RC) model to obtain the approximation \( \mathbf{\Theta}(\cdot ;  \cdot) \);
\vspace{2mm}
    \STATE \textbf{Online Stage:}
    \vspace{1mm}
    \STATE Evaluate  the output \( \mathbf{\Theta}(t^* ; \boldsymbol{\mu}^*) \) of the RC model for a new time-parameter pair \( (t^*, \boldsymbol{\mu}^*) \);
    \vspace{1mm}
    \STATE Compute the reduced-order solution:  
    \(
    C_{rb}\left(t^*; \boldsymbol{\mu}^{*}\right) = \mathcal{V} \mathbf{\Theta}\left(t^{*}; \boldsymbol{\mu}^{*} \right); 
    \)
\end{algorithmic}
\end{algorithm}

Incorporating RC architectures into ROM offers several advantages. One key benefit is the reduction in training costs, as only the output layer is optimized, eliminating the need for backpropagation. Additionally, RC effectively captures temporal features by mapping input data into a high-dimensional space via the reservoir. Specifically, ESNs, a widely used approach within RC, ensure long-term stability due to the echo state property, making them highly efficient for modelling complex dynamical systems. Furthermore, RC can achieve the same level of learning accuracy with less training data and computational resources compared to conventional deep learning architectures, such as RNNs, LSTM networks, and DNNs. In this work, a reservoir computing architectures library \text{reservoirPy} (\href{https://github.com/reservoirpy/reservoirpy}{{https://github.com/reservoirpy/reservoirpy}}) \cite{Trouvain2020} is employed to implement RC within the ROM framework.  \\

{{To assess the accuracy of the ROM predictions for steel grade intermixing, we evaluate the error on the quantity of interest (QoI), defined as the concentration of the new steel grade measured at the tundish outlet (via locating a probe), which directly characterizes the temporal evolution of steel grade intermixing. The relative $L_2$ error between the ROM prediction \( C_{rb}(t; \boldsymbol{\mu}) \) and the high-fidelity solution \( C(t; \boldsymbol{\mu}) \) is given by
\begin{equation}
\mathcal{L}_{2} = 
\frac{\| C - C_{rb} \|_2}{\| C \|_2} 
= \frac{\sqrt{\sum_{i=1}^{N_t} \left( C( \boldsymbol{\mu}, t_i) - C_{rb}(\boldsymbol{\mu}, t_i) \right)^2}}{\sqrt{\sum_{i=1}^{N_t} C(\boldsymbol{\mu}, t_i)^2}}.
\label{eqn:l2_error}
\end{equation} }}

%% file: sections/domain.tex
\section{Numerical Setup}\label{sec:num_setup}

This section outlines the numerical setup employed for the full-order model simulations. The geometry and mesh characteristics of the tundish system are presented in \ref{subsec:computational_domain}. The tundish operating and boundary conditions considered for the transient ladle changeover operation are reported in \ref{BCs}.

\subsection{Computational domain}\label{subsec:computational_domain}

\begin{figure}[ht!]
    \centering
    \includegraphics[width=1\linewidth]{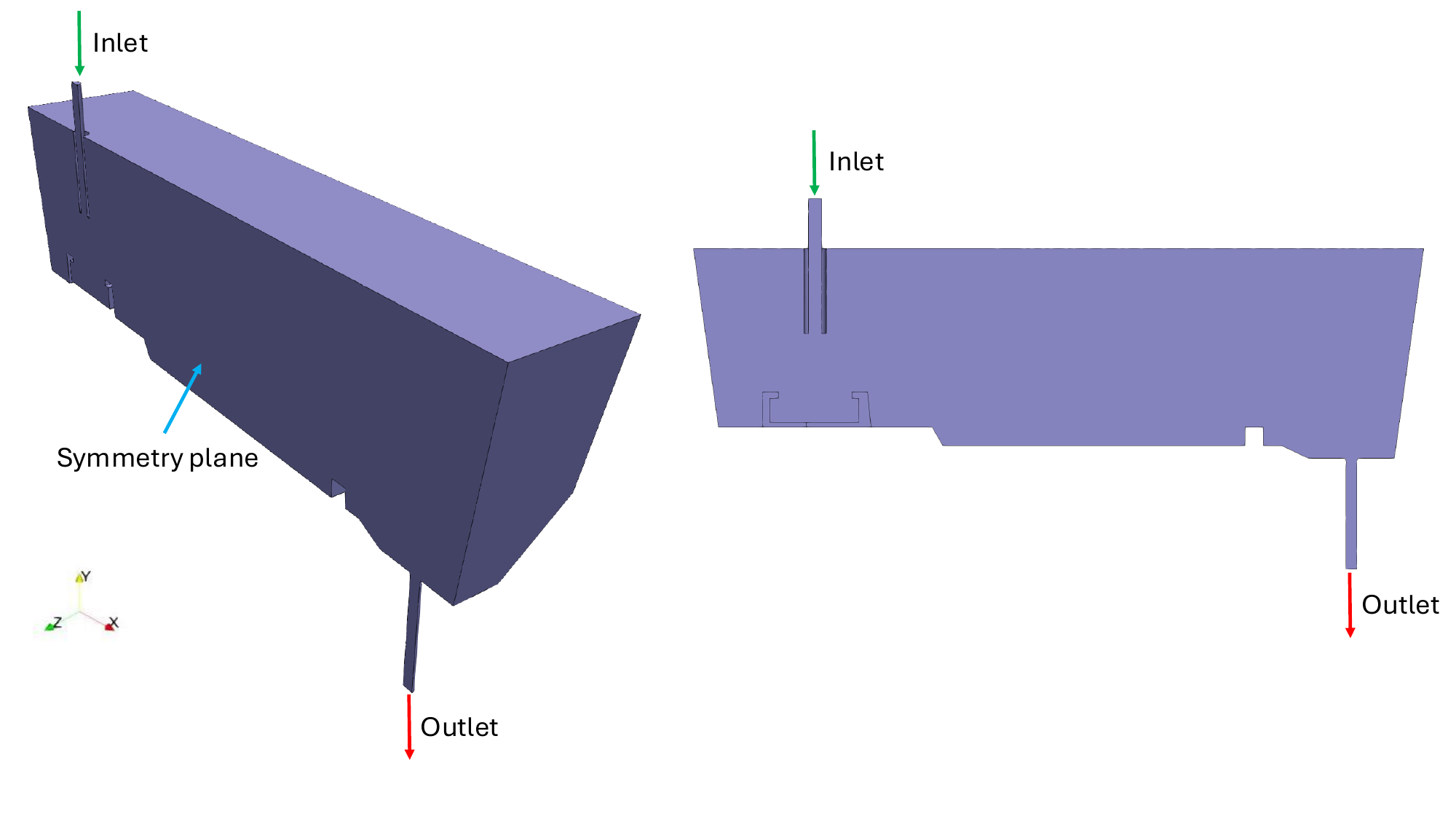}
    \caption{Computational domain of a 0.5 scaled single strand tundish. \textbf{Left}: Isometric view highlighting the inlet, outlet, and symmetry plane, with all other boundaries treated as walls. \textbf{Right}: 2D left-side view overlooking the symmetry plane.}
    \label{fig:Geometry}
\end{figure}

{{{A three-dimensional computational domain is developed based on a 0.5 scaled industrial tundish, as shown in Figure \ref{fig:Geometry}. The industrial tundish has a nominal capacity of 70~ton (10.0~m$^3$) with a bottom plan dimension of 1330~mm $\times$ 5550~mm. The side walls are inclined at 8$^\circ$, with a 45$^\circ$ chamfer along the long-side bottom edge, and the centre-to-centre distance between the ladle shroud axis and the outlet axis is 4400~mm. The tundish bottom incorporates two steps to reduce residual steel at the end of casting. \\

The internal flow control devices include a rectangular impact pot (1000~mm $\times$ 890~mm $\times$ 350~mm, full scale), centrally positioned with respect to the ladle shroud axis, and a dam located 3500~mm downstream of the shroud axis with a height of 150~mm, which prevents premature discharge of cold steel during the initial filling stage. \\

Geometric similarity is preserved by uniformly scaling all dimensions by a factor of 0.5, resulting in a scaled liquid steel volume of 1.25~m$^3$. To reduce computational cost, only half of the domain is considered by exploiting symmetry with respect to the longitudinal plane, yielding a computational domain volume of 0.625~m$^3$. All operating conditions correspond to a 1:2 physical model and are scaled using the Froude similarity criterion to ensure dynamic similarity between model and prototype. \\

The nominal steel bath level at 70~ton capacity is 1210~mm (full-scale), measured from the tundish bottom at the impact zone, which is adopted as the reference datum due to the stepped geometry. The ladle shroud is submerged 200~mm below the free surface to ensure stable jet formation at the inlet.}}} \\

\begin{figure}[ht!]
    \centering
    \includegraphics[width=1\linewidth]{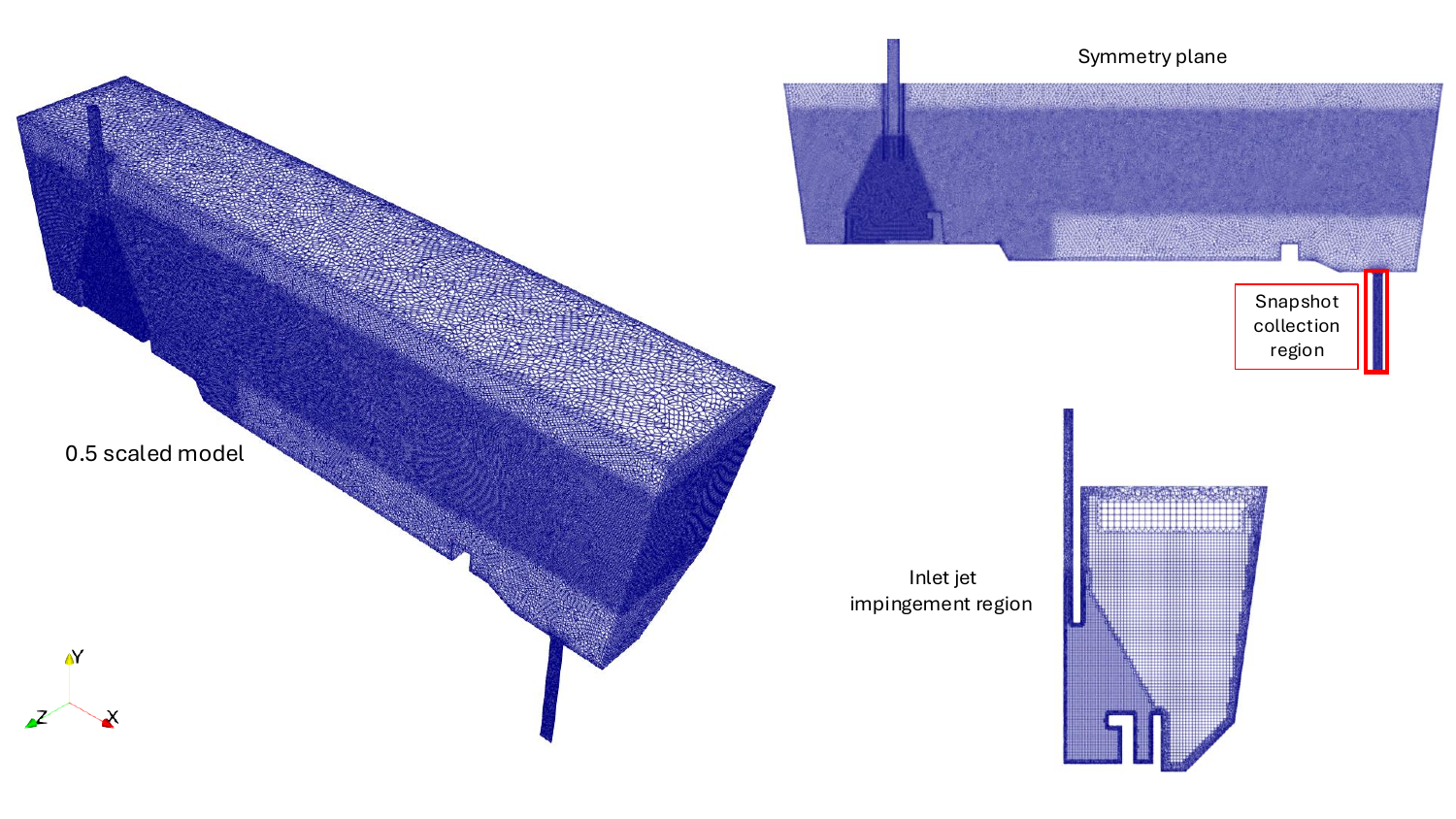}
    \caption{Discretized computational domain of a 0.5-scale single-strand tundish. \textbf{Left}: Isometric view of discretized domain; \textbf{Top right}: 2D symmetry plane showing the mesh refinement strategies to capture accurately the interface development and turbulence mixing phenomena. The entire outlet region, highlighted by the red box, is used for snapshot collection; \textbf{Bottom right}: Inner jet impingement zone with an impact pot to enhance turbulence mixing.}
    \label{fig:Mesh_ladle_change}
\end{figure}

The computational mesh consists of 2,181,611 cells, consisting of 1,496,381 hexahedral and 685,230 polyhedral elements, as shown in Figure \ref{fig:Mesh_ladle_change}. To enhance near-wall resolution, three prism layers are incorporated. The mesh is refined in the region of interface movement to accurately capture phase interactions. Additionally, further refinement is applied in the inlet jet impingement zone, as illustrated in the bottom right of Figure \ref{fig:Mesh_ladle_change}, to accurately capture the turbulent mixing phenomena. {{{A refined mesh is also employed in the outlet duct region, where grade intermixing is evaluated and concentration snapshots are extracted, ensuring adequate resolution of velocity and concentration gradients in this zone.}}}

\subsection{Tundish operating and boundary conditions}\label{BCs}

{{The liquid steel and slag phases are modelled as incompressible Newtonian fluids with constant density and dynamic viscosity, as reported in Table~\ref{tab:properties} and the properties of the two steel grades are assumed to be identical.}} Table \ref{Table:tundish_operating_parameters} presents the operating parameters of the tundish. Table \ref{Table:IntialBCs} summarizes the boundary conditions applied during the three significant stages of the ladle changeover operation of the tundish.{{ At the inlet, the hydraulic diameter is set equal to the ladle shroud diameter. The turbulence intensity was computed using the standard empirical relation, is defined as
\begin{equation*}
    I = 0.16 \, Re_{D_h}^{-1/8}.
\end{equation*}
During the ladle changeover (refilling stage), the turbulence intensity is updated consistently with the varying inlet flow rate to ensure accurate representation of transient flow conditions.}
\begin{table}[ht!]
\centering
\renewcommand{\arraystretch}{1.2}
\begin{tabular}{|c|c|c|}
\hline
Phase & Density $\rho$ [kg/m$^{3}$] & Dynamic viscosity $\mu$ [kg/(m\,s)] \\
\hline
Liquid steel & 7000 & $5.55 \times 10^{-3}$ \\
Slag         & 2560 & $0.6$ \\
\hline
\end{tabular}
\caption{Thermophysical properties of the liquid steel and slag phases used in the simulations.}
\label{tab:properties}
\end{table}}
\begin{table}[ht!]
    \centering
    \renewcommand{\arraystretch}{1.2}
    \begin{tabular}{|>{\centering\arraybackslash}p{6cm}|>{\centering\arraybackslash}p{3cm}|}
        \hline
        \textbf{Tundish operating parameters} & \textbf{Value} \\
        \hline
        Height of molten steel (mm) & 458.5 \\
        Inlet nozzle diameter (mm) & 55 \\ 
        Outlet nozzle diameter (mm) & 45 \\
        Tundish throughput (m$^3$/hr) & 9.77 \\ 
        Inlet throughput velocity (m/s) & 1.1424 \\
        Outlet throughput velocity (m/s) & 1.7069 \\
        Filling flow rate (m$^3$/hr) & 25.2538 \\
        Filling period inlet velocity (m/s) & 2.9563 \\
        \hline
    \end{tabular}
    \caption{Operating conditions of the tundish}
    \label{Table:tundish_operating_parameters}
\end{table}
\begin{center}
\begin{table}[ht!]
    \centering
    \renewcommand{\arraystretch}{1.2}
    \small 
    \begin{tabular}{|l|l|} 
        \hline
        \multicolumn{2}{|c|}{\textbf{Stage 1: Steady-state operation}} \\
        \multicolumn{2}{|c|}{\textbf{(RANS steady and URANS, VOF - two phase flow model)}} \\
        \hline
        Inlet  & Constant velocity at inlet, volume fraction of old grade throughout = 1.0 \\
        & velocity-based turbulence kinetic energy and dissipation rate \\
        Outlet & flowRateOutletVelocity condition to maintain a constant outflow rate. \\
        Walls  & No slip with standard wall function.\\
        \hline
        \multicolumn{2}{|c|}{\textbf{Stage 2: Tundish emptying operation}} \\
        \multicolumn{2}{|c|}{\textbf{( URANS, VOF - two phase flow model)}} \\
        \hline
        Inlet  & No inflow. \\
        Outlet & flowRateOutletVelocity condition to maintain a constant outflow rate. \\
        Walls  & No slip with standard wall function. \\
        \hline
        \multicolumn{2}{|c|}{\textbf{Stage 3: Tundish refilling operation}} \\
        \multicolumn{2}{|c|}{\textbf{(URANS, VOF - two phase flow model + scalar transport equation)}} \\
        \hline
        Inlet  & Fixed or varying desired inlet velocity with volume fraction of new grade = 1.0 \\ 
               & velocity-based turbulence kinetic energy and dissipation rate \\
        Outlet & flowRateOutletVelocity condition to maintain a constant outflow rate. \\
        Walls  & No slip with standard wall function; \\ 
        \hline
    \end{tabular}
\caption{Initial and boundary conditions considered for the three different stages of ladle changeover operations of a single-strand tundish.}
\label{Table:IntialBCs}
\end{table}
\end{center}

{{{To simplify the numerical model, as mentioned earlier only two phases are considered in the simulations, molten steel and slag. The surrounding atmosphere (air) is neglected, and the entire volume above the steel bath is assumed to be occupied by slag. A pressure-outlet boundary condition is imposed at the top surface, with the backflow phase fraction specified as 100 $\%$ slag. During ladle changeover, as the steel level decreases, reverse flow through the top boundary allows slag to enter the computational domain and compensate for the volume reduction. Consequently, the slag layer thickness varies during the transient simulation, unlike in real plant operation where the slag thickness remains approximately constant and the volume above it is occupied by air. Although a three-phase model including air would provide a more realistic representation of the free surface behaviour, the present simplification is adopted to improve numerical stability and reduce computational complexity. Since the primary objective of the study is to investigate steel grade intermixing within the tundish bulk flow, this approximation is not expected to significantly affect the predicted transition behaviour.}}}

%% file: sections/numerical_method.tex
\section{Numerical method}\label{sec:num_method}

Among the numerical methods used for multiphase flows, the open-source VOF solver \text{interFoam} has gained significant attention and widespread adoption \cite{Deshpande2012}. Initially developed by Ubbink \cite{Ubbink_1997} within the FOAM framework \cite{Jasak2009}, \text{interFoam} has undergone continuous modifications and improvements. It is now part of the OpenFOAM suite of C++ libraries, designed for the finite-volume discretisation of partial differential equations, particularly related to continuum mechanics. The object-oriented structure of C++ enables the code to closely mirror its mathematical formulation, while also making the high-level syntax flexible for further development and modification \cite{Weller1998}. \\

In this study, we use the \text{interFoam} solver from OpenFOAM, a two-phase incompressible, isothermal and immiscible fluid solver that employs a VOF phase fraction based approach for interface capturing. It also supports optional mesh motion and topology modifications, including adaptive re-meshing, and we highlight that these strategies were not employed in this study. For a detailed description of the solver, the reader is referred to \cite{Deshpande2012}. In this work, the momentum equation is solved using the PIMPLE algorithm, which combines the PISO and SIMPLE methods to improve stability and convergence. The solution procedure involves computing an initial velocity field, followed by multiple pressure correction loops to ensure accuracy. \\

\begin{table}[ht!]
    \centering
    \renewcommand{\arraystretch}{1.2}
    \begin{tabular}{|>{\centering\arraybackslash}p{7cm}|>{\centering\arraybackslash}p{5cm}|}
        \hline
        \textbf{Term} & \textbf{Scheme} \\
        \hline
        Time derivative & Euler \\
        Convective term (momentum) & linearUpwind \\
        Convective term (volume fraction \(\alpha\)) & MULES (vanLeer + interface compression) \\
        Convective term (species: \(C\)) & upwind \\
        Convective term (turbulence: \(k\), \(\varepsilon\)) & upwind \\
        Diffusive term (viscous) & linear \\
        Gradient term & cellMDLimited Gauss \\ 
        \hline
    \end{tabular}
    \caption{Discretization schemes employed in the interFoam based multiphase species transport solver for ladle changeover operation transient simulation.}
    \label{table:discretization_scheme}
\end{table}

Table \ref{table:discretization_scheme} summarizes the discretization schemes employed for solving the governing equations. For the volume fraction $\alpha$ and species concentration \textit{C}, the Multi-dimensional Universal Limiter for Explicit Solution (MULES) scheme is utilized, specifically designed for solving equations where the solution variable is subject to lower and upper bounds. Its primary objective is to ensure that the computed solution remains within these bounds while maintaining accuracy. Originally developed for interface capturing in the VOF method, where it enforces phase fraction $\alpha$ to remain within its physical limits of 0 and 1. Second-order linear upwind schemes are used for the convective terms in both the momentum and volume fraction advection equations. First-order upwind schemes are used for other terms, and the Gauss method is employed for gradient computation. Appropriate under-relaxation was applied to ensure numerical stability and improve the convergence rate of the iterative solution process, with optimal parameters determined through trial-and-error. {{Additionally, the time step is selected to maintain a maximum Courant number of 2, ensuring stable and converged solutions, as the present study focuses on bulk intermixing dynamics rather than detailed steel–slag interface resolution, with the VOF method employed solely to track the liquid steel level.}} \\

Considering these tundish operating parameters, the specified initial and boundary conditions, the following methodology is employed to model the three distinct stages of ladle changeover and grade intermixing in the tundish: 
\begin{enumerate}[leftmargin=*, label=\textbf{Stage \arabic*}:]
    \item \textbf{Steady-state operation} – A steady-state simulation is first conducted to obtain the converged velocity and turbulence fields, which serve as initial conditions for the subsequent VOF simulation with a tundish filled with molten steel and slag. The pre-ladle changeover phase is then modelled using the VOF method, treating it primarily as a turbulent, homogeneous flow problem. The computed flow and turbulence fields within the active tundish volume are stored to initialize the subsequent transient simulation.
    
    \item \textbf{Transient emptying operation} – The tundish emptying phase, occurring without inflow, is modelled using a transient, two-phase VOF approach, considering molten steel and slag as immiscible fluids. The flow and turbulence fields from Stage 1 serve as initial conditions, while the mass inflow is set to zero with constant outflow. Within this framework, the evolution of mixture velocity, turbulence and phase volume fraction in the active tundish volume is computed over time.

    \item \textbf{Transient refilling operation} – The refilling and grade intermixing phase is simulated using a numerical procedure similar to that described in Stage 1. In this stage, the velocity and turbulence fields, along with the volume fraction distribution obtained from Stage 2, are used as initial conditions for the simulation.
\end{enumerate}

During all three stages, the tundish outflow is maintained at a constant predetermined rate to replicate the steady casting conditions characteristic of industrial continuous casting processes.

%% file: sections/results.tex
\section{Results and discussion}\label{sec:result}

In this section, we present and analyze the results obtained from the FOM and the ROM developed to evaluate steel grade intermixing during sequential casting. The performance of the FOM is first examined, with particular emphasis on the influence of ladle change time on flow dynamics and species mixing. Subsequently, the ROM predictions are evaluated and compared against the FOM results to assess the accuracy and efficiency of the reduced modelling framework.

\subsection{FOM results}

\begin{figure}[ht!]
    \centering
    \includegraphics[width=1\linewidth]{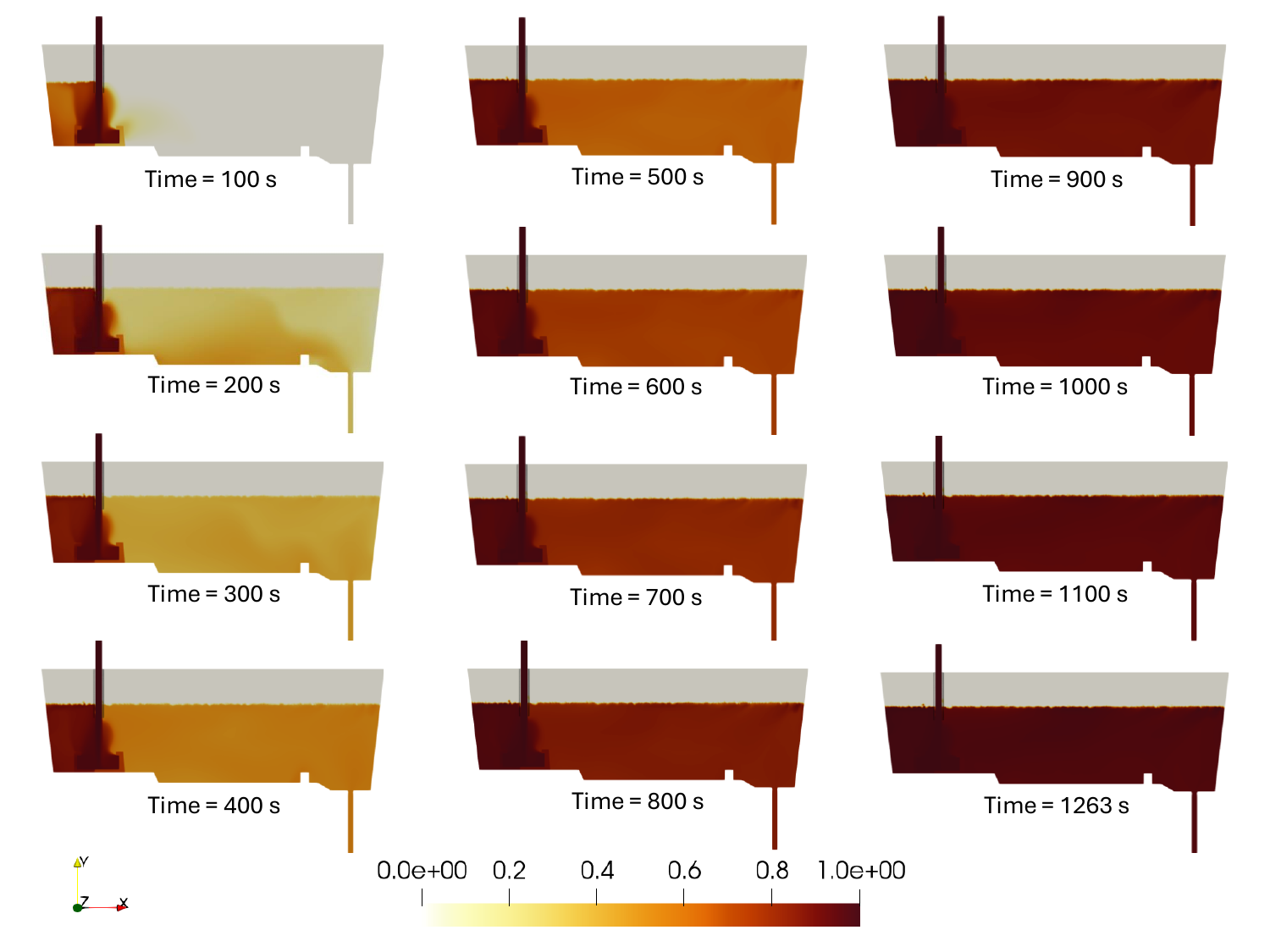}
    \caption{New steel grade transition contours corresponding to a ladle change duration of 64s and a filling time of 40.40s.}
    \label{fig:SteelGradeTransitionContour}
\end{figure}

\begin{figure}[ht!]
    \centering
    \includegraphics[width=0.65\linewidth]{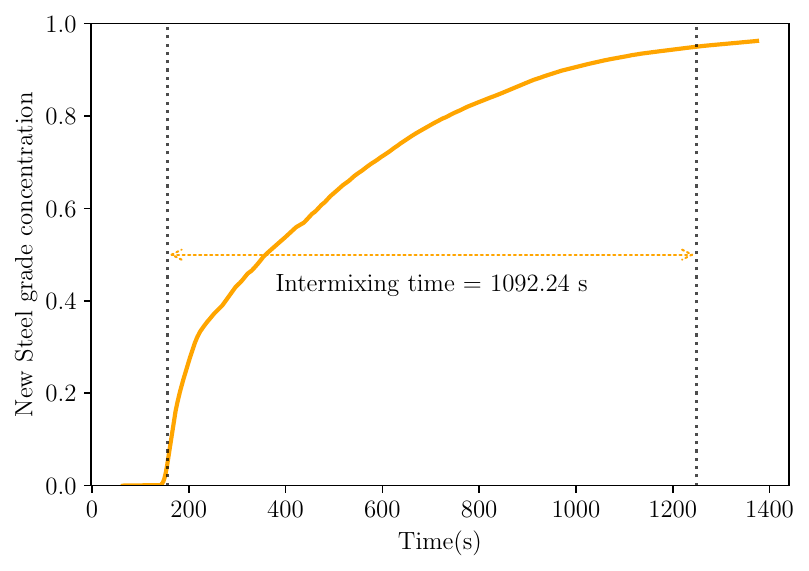}
    \caption{Intermixing time corresponding to a ladle change duration of 64s and a filling time of 40.40s.}    \label{fig:initial_IMT}
\end{figure}

For the initial parameter study, we consider the tundish throughput rate 9.77 $m^3/hr$ with a ladle change time of 64 s and corresponding filling duration of 40.40 s, with a filling flow rate of 25.25 $m^3/hr$ for the newly replaced ladle. During the filling period, the volume of steel emptied during the ladle change process is replaced by the new steel grade. Figure \ref{fig:SteelGradeTransitionContour} shows the contours of the new steel grade evolution. A probe is located at the tundish outlet to monitor the new steel grade evolution. The new steel concentration recorded by the probe positioned at the outlet is shown in Figure \ref{fig:initial_IMT}. For this initial parameter, ladle change time of 64 s, the intermixing time determined by the FOM is 1092.24  s.  

\subsubsection{Effects of ladle change time} 

In this study, we examine the impact of ladle change time on the intermixing behaviour of a new steel grade in a single strand tundish. Five different ladle change times and their corresponding filling durations, required to restore the melt height to steady-state levels, are considered. Table \ref{table:effectorLCTonIMT} summarizes the parameters used in the numerical experiments. {{{The chosen parameter range (64 s -- 305 s in the scaled model) is constrained by industrial operating limits associated with maintaining a sufficient steel head in the tundish during ladle changeover. The lower bound corresponds to the minimum feasible ladle change duration, while the upper bound represents the maximum allowable draining time before critical operating conditions are approached. This ensures safe operation by maintaining an adequate steel head required for stable outflow conditions and to avoid adverse flow phenomena in the tundish.}}} Here, the primary quantity of interest (QoI) is the {{intermixing time, defined as the time required for the concentration of new grade steel to increase from 5$\%$ to 95$\%$ at the tundish outlet.}} \\

During the ladle change operation cycle, the tundish melt bath height remains low when a new ladle is introduced to supply molten metal. To rapidly restore steady-state melt height, the inflow rate from the ladle shroud is typically increased. At the start of pouring the new steel grade, a substantial volume of residual steel from the previous grade remains in the tundish, referred to as the residual steel volume (RVF). As the ladle change time increases, the filling duration also extends to maintain the melt height. Since ladle change time directly influences RVF, it has a significant impact on the intermixing of the two steel grades. \\

\begin{table}[ht!]
\centering
\begin{tabular}{|c|c|c|}
\hline
\textbf{Expt. No} & \textbf{Ladle change time (s)} & \textbf{Filling time (s)} \\
\hline
1 & 64 & 40.40  \\
2 & 125 & 78.92  \\
3 & 185 & 116.79  \\
4 & 245 & 154.67 \\
5 & 305 & 192.55 \\
\hline
\end{tabular}
\caption{Numerical experiment parameters considered to investigate the effects of ladle change time, ladle change time with corresponding filling time to reach steady-state operation.}
\label{table:effectorLCTonIMT}
\end{table}

\begin{figure}[ht!]
    \centering
    \includegraphics[width=0.9\linewidth]{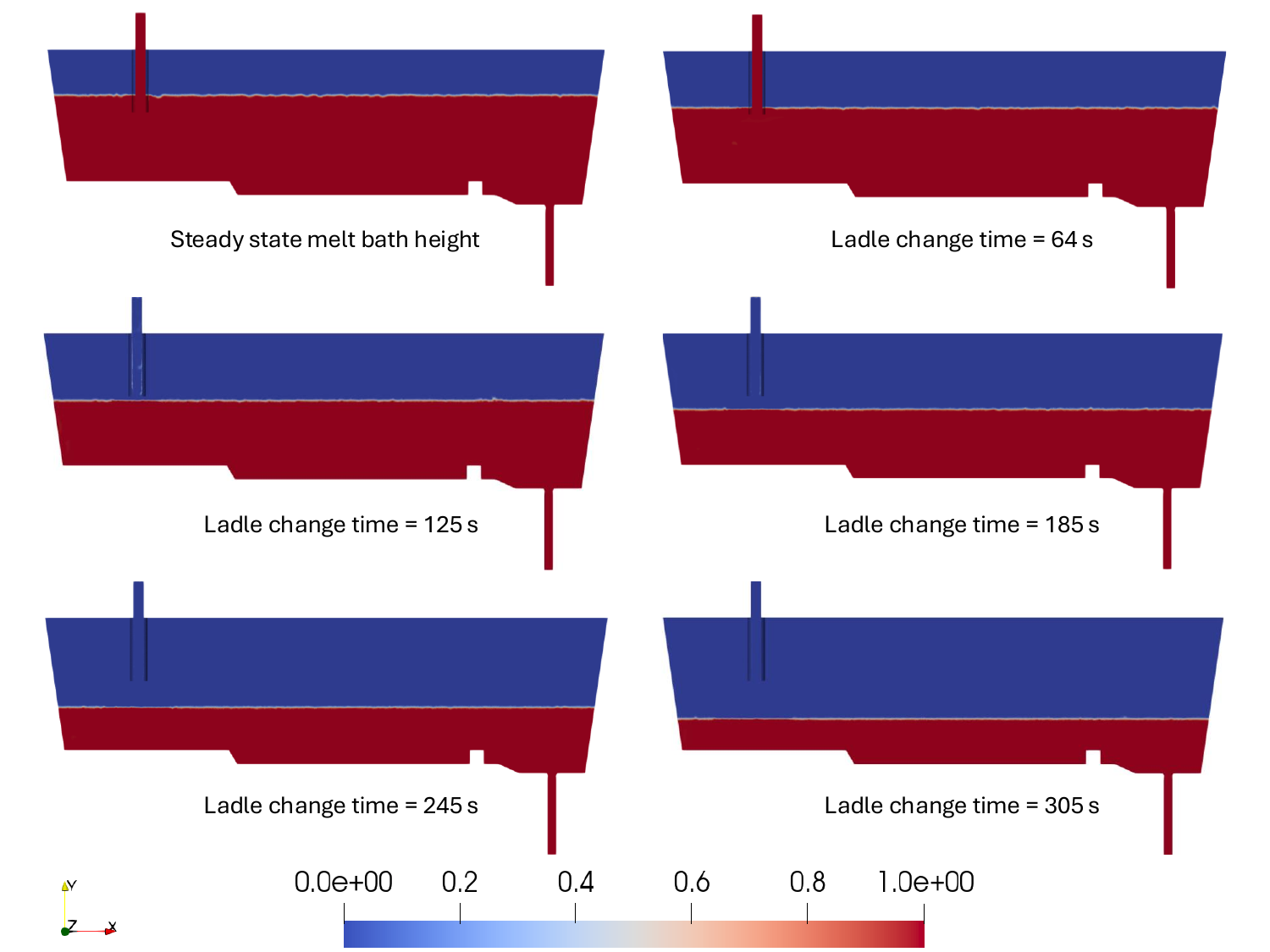}
    \caption{Effect of ladle change time on melt bath height and residual volume fraction.}
    \label{fig:RVF}
\end{figure}

\begin{figure}[ht!]
    \centering
    \includegraphics[width=0.65\linewidth]{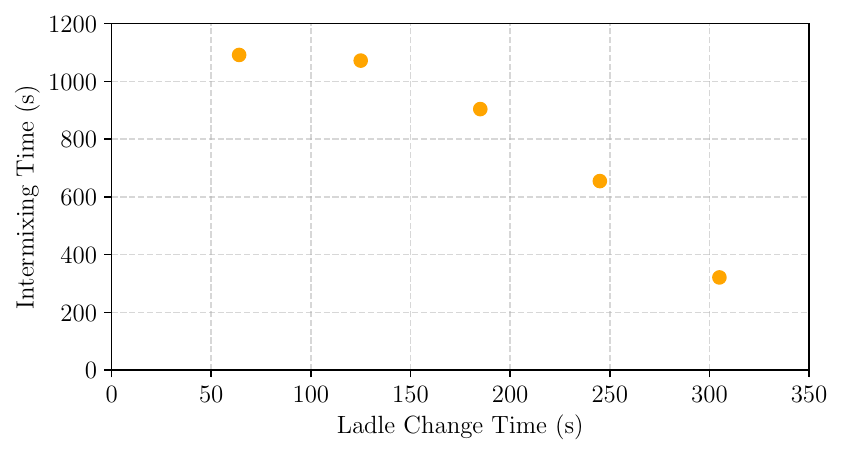}
    \caption{Effect of ladle change time on intermixing time}
    \label{fig:LCTeffect}
\end{figure}

The effect of ladle change time on melt bath height and residual steel volume (RVF) is illustrated in Figure \ref{fig:RVF}. The results indicate that as ladle change time increases, the melt bath height decreases, and similarly, the RVF decreases. The mixing phenomena of two different miscible steel grades are influenced by variations in viscosity and density. While macroscopic flow dynamics primarily govern the mixing behaviour, microscopic interactions also contribute to the process. The extent of mixing is largely determined by turbulent kinetic energy, which facilitates the exchange of kinetic energy between the mean flow and fluctuating layers of fluids \cite{Siddiqui2015, Siddiqui2019}. Figure \ref{fig:LCTeffect} shows the effect of ladle change time on the intermixing time. As illustrated, the intermixing time decreases with increasing ladle change time. It reveals that the intermixing time was shortest for the lowest residual volume and longest for the highest residual volume in the tundish. This trend can be attributed to the reduction in RVF, which leads to more efficient mixing of the new and old steel grades in the initial stages. It indicates that longer ladle change times promote faster intermixing of the two steel grades, thereby accelerating the homogenization of the new steel grade within the tundish. This study is essential for optimizing the steel production process.

\subsection{ROM results}

\textit{Database creation}: Prior to applying the model order reduction strategies, it is necessary to prepare a database consisting of full-order solutions. Since the primary focus is on the outlet to obtain the QoI, the snapshot database is constructed by considering only the outlet region. We start collecting snapshots where the concentration of the new steel grade is 1$\%$ at the outlet. A database is then generated, consisting of five numerical experiment snapshots of the new steel grade, $C(t_k, \boldsymbol{\mu}_k)$, obtained from the FOM. We consider the first four numerical experiments as a training set and the last one as a test set. {{{Flow field snapshots are extracted at uniform intervals of 1 s throughout each simulation. The resulting snapshots are assembled into a global snapshot matrix of size $N_h \times N_s$ (i.e., $68,509 \times 3890$), where $N_h$ denotes the number of spatial degrees of freedom (at outlet region) and $N_s$ the total number of snapshots collected from the training set.}}} \\

\begin{figure}[ht!]
    \centering
    \includegraphics[width=0.8\linewidth]{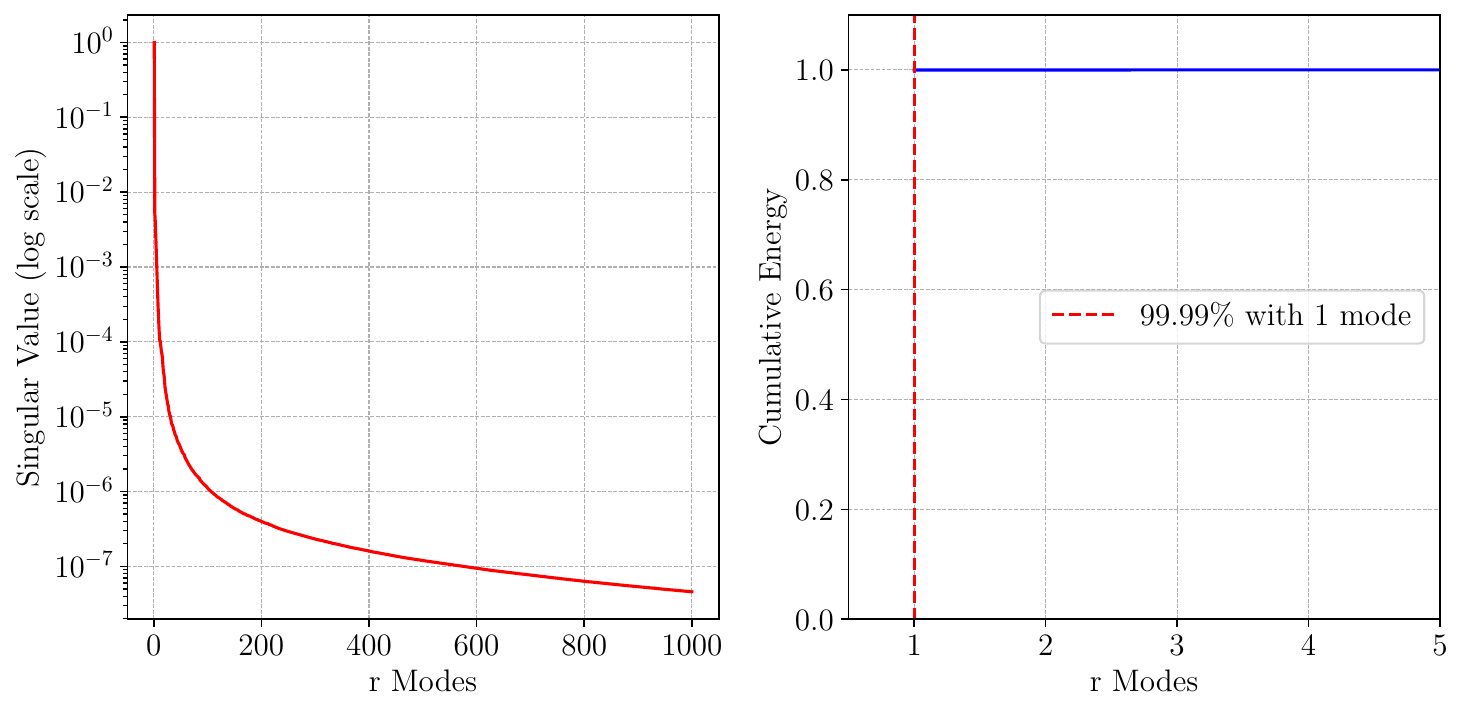}
    \caption{Singular values and cumulative energy distribution of the training dataset used in developing the POD-RC-ROM for predicting intermixing time.}
    \label{fig:singular_values}
\end{figure}

Further, we perform model reduction using POD on the training set snapshots to obtain the reduced basis subspace. Figure \ref{fig:singular_values} shows both singular values and cumulative energy corresponding to the rank of POD modes. We notice a sharp decay in singular values, with the first mode alone capturing 99.99 $\%$ of the cumulative energy. This rapid decay enables the construction of an efficient ROM using a low-dimensional linear approximation subspace. \\

We utilize the first POD mode to construct a reduced basis subspace and then follow the procedure outlined in section \ref{subsec:POD} to obtain modal coefficients via Galerkin projection. To enable rapid online computations, reservoir computing is employed to establish a mapping between the parameters and modal coefficients. This approach facilitates the development of a data-driven POD-RC based ROM for monitoring steel grade intermixing time. \\

\begin{figure}[ht!]
    \centering
    \includegraphics[width=0.65\linewidth]{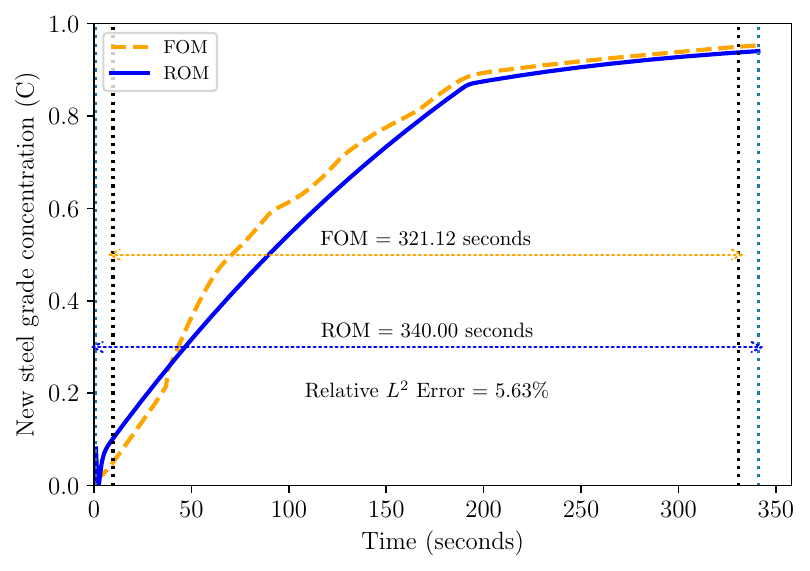}
    \caption{Prediction of intermixing time by the POD-RC-ROM for test set parameter in the extrapolation regime.}
    \label{fig:prediction}
\end{figure}

Figure \ref{fig:prediction} presents the prediction obtained from the POD-RC-ROM for the test parameter, compared against the FOM results, and additionally, it reports the relative $L_2$ error of the prediction, which is 5.63$\%$. The results demonstrate that the ROM predictions are both qualitatively and quantitatively accurate, closely following the trend of the FOM results. \\

\begin{figure}[ht!]
    \centering
    \includegraphics[width=0.65\linewidth]{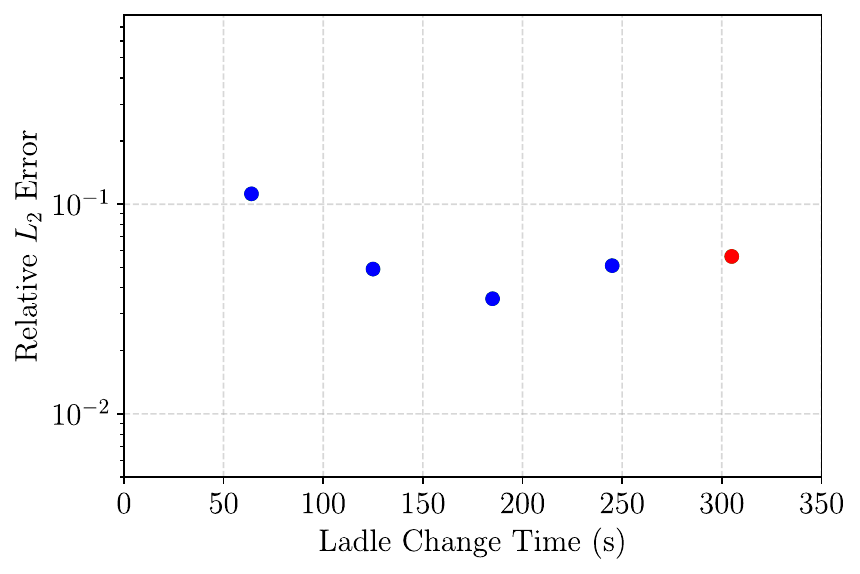}
    \caption{Prediction relative $L_2$ error of POD-RC-ROM for training and test sets (blue and red, respectively).}
    \label{fig:predictionerror}
\end{figure}

The relative $L_2$ error of ROM predictions for both the training and test set parameters is shown in Figure \ref{fig:predictionerror}. The observed decrease in error for the training cases suggests that the ROM is progressively learning the system dynamics and improving its approximation capability. The reduction in error indicates that the reservoir computing framework successfully captures the dynamics of the system, leading to enhanced predictive accuracy. For the test case, even though there's a slight increase in relative $L_2$ error, the ROM provide reliable results as shown in Figure \ref{fig:prediction}. \\

The decreasing trend in training errors confirms the effectiveness of the reservoir computing-based ROM in learning the reduced order dynamics. The marginal increase in the test set error suggests that further refinements, such as incorporating additional training samples or adjusting hyper-parameters could enhance the generalization capability of the model. \\

\begin{figure}[ht!]
    \centering
    \begin{subfigure}[b]{0.65\linewidth}
        \centering
        \includegraphics[width=\linewidth]{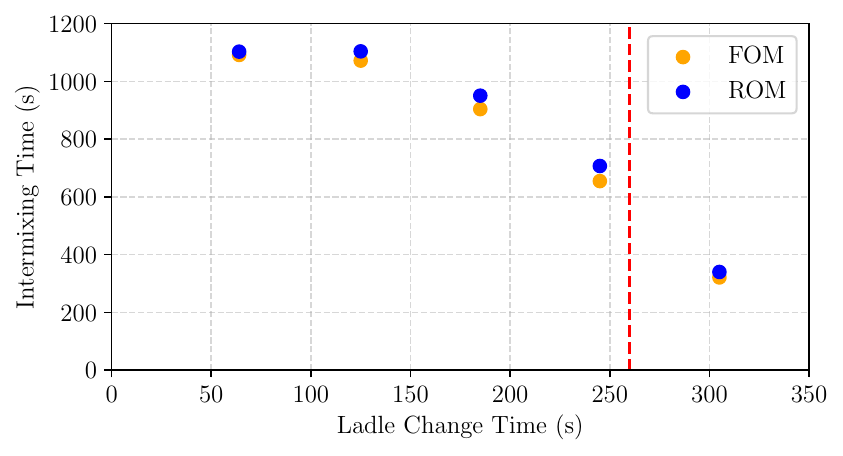}
        \label{fig:scatter}
    \end{subfigure}
    \begin{subfigure}[b]{0.65\linewidth}
        \centering
        \includegraphics[width=\linewidth]{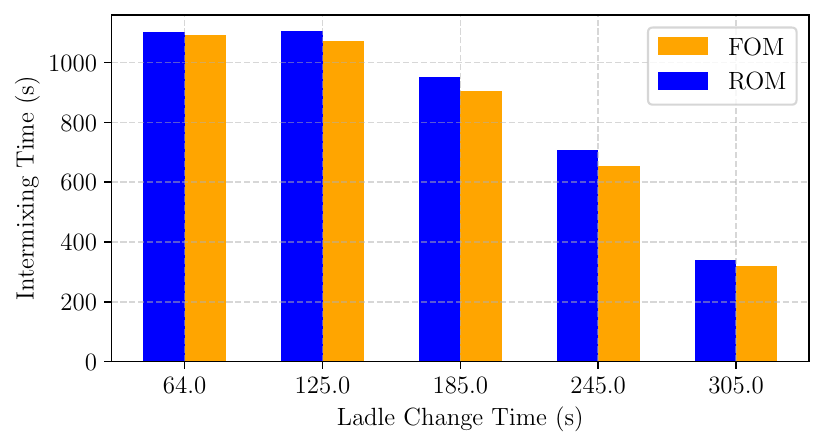}
        \label{fig:bar}
    \end{subfigure}
    \caption{Comparison of intermixing time obtained from the FOM and the POD-RC based ROM. The red vertical line separates the training and test sets, with the left region representing the training set and the right region representing the test set. }
    \label{fig:IntermixingTimeFOM-ROM-prediction}
\end{figure}

We use the resulting POD-RC-ROM to predict intermixing time. Figure \ref{fig:IntermixingTimeFOM-ROM-prediction} compares the intermixing time obtained from the FOM and ROM. The results show that the ROM predictions closely align with the FOM values for the train and test parameters, with a negligible difference compared to the FOM. \\

\begin{table}[ht!]
    \centering
    \renewcommand{\arraystretch}{1.2} 
    \begin{tabular}{|>{\centering\arraybackslash}p{1.7cm}|>{\centering\arraybackslash}p{4.5cm}|>{\centering\arraybackslash}p{3.2cm}|>{\centering\arraybackslash}p{2.8cm}|}        
    \hline
    \textbf{Expt No.} & \textbf{Wall-clock time} & \textbf{CPU time (core-hours)} & \textbf{Cores Allocated} \\     
    \hline
    1 &  351.6 hours (14.7 days) &  78,810 hours    &   200 \\ 
    2 &  527.6 hours (22.0 days) &  105,318 hours   &   200 \\ 
    3 &  396.5 hours (16.5 days) &  73,535 hours   &   192 \\ 
    4 &  335.7 hours (14.0 days) &  64,176 hours   &   192 \\  
    5 &  161.83 hours (6.74 days)  &  38,203 hours   &   256 \\  
    \hline
    \end{tabular}
    \caption{Summary of computational resources used for the ladle changeover FOM transient simulations on the Linux cluster. Wall-clock time refers to the elapsed real time, and CPU time represents the cumulative time across all allocated cores (core-hours). These experiments correspond to those reported in Table \ref{table:effectorLCTonIMT}.}
    \label{tab:LCT_FOMcomp-resources}
\end{table}

\begin{table}[ht!]
    \centering
    \renewcommand{\arraystretch}{1.2} 
    \begin{tabular}{|>{\centering\arraybackslash}p{3cm}|
                    >{\centering\arraybackslash}p{3.2cm}|
                    >{\centering\arraybackslash}p{3.2cm}|
                    >{\centering\arraybackslash}p{2.5cm}|}
    \hline
    \textbf{Data-driven ROM} & \textbf{Offline Time (Wall-clock)} & \textbf{Online Time (Wall-clock)} & \textbf{Cores Used (Offline / Online)} \\
    \hline
    POD-RC-ROM &  75 seconds &  0.42 seconds & 1 / 1 \\
    \hline
    \end{tabular}
    \caption{Computational resources used for the ladle changeover operation, ROM construction, and prediction. The offline phase includes the computation of POD modes (only the outlet region of the domain is considered), evaluation of modal coefficients, and the training of the reservoir computing model to construct a regression map. The online phase corresponds to the prediction of the quantities of interest at the tundish outlet.}
    \label{tab:LCT_ROMcomp-resources}
\end{table}

The computational resources and wall-clock time required for the FOM simulations of the ladle changeover process are summarised in Table~\ref{tab:LCT_FOMcomp-resources}. These high-fidelity simulations exhibit substantial computational demand, with wall-clock times ranging from approximately 161.8 to 527.6 hours (6.74 to 22.0 days), executed on 192 to 256 cores. The associated CPU times span from 38,203 to 105,318 core-hours. In contrast, the performance of the data-driven ROM, based on POD and RC, is reported in Table~\ref{tab:LCT_ROMcomp-resources}. The offline training stage, which includes the computation of POD modes, modal coefficient computation, and RC based regression mapping, requires 75 seconds of wall-clock time on a single core. The ROM’s online prediction of the QoIs at the tundish outlet is completed in 0.42 seconds, also on a single core. Although the offline cost is non-negligible, it is incurred only once and enables rapid repeated evaluations thereafter. The ROM achieves a reduction in wall-clock time exceeding five orders of magnitude relative to the FOM, indicating its potential suitability for time-sensitive applications such as real-time monitoring, control, and optimization.

%% file: sections/conclusion.tex
\section{Conclusion and perspectives}\label{sec:conslusion}

To ensure continuous casting and enhance productivity, different steel grades are cast sequentially through a ladle changeover operation in the continuous casting process. The key objective is to minimize grade intermixing, as the resulting intermixed region is typically regarded as inferior quality and often downgraded. A three-dimensional, transient, two-phase turbulent flow has been carried out to investigate ladle changeover operation and the associated grade intermixing within the tundish. The effect of ladle change time on intermixing time is studied, revealing that ladle change time significantly influences the melt height and residual volume fraction (RVF). Moreover, RVF plays a crucial role in grade intermixing within the tundish. A lower RVF facilitates faster intermixing of grades, while a higher RVF tends to prolong the intermixing duration. Generally, for industrial-scale systems, obtaining full-order high-fidelity solutions of these physical processes is computationally expensive and requires long-time integration to simulate the entire operations cycle. This makes it prohibitive to perform extensive parameter exploration, which is crucial for the development of ROM. \\

To address this issue, a POD-RC based ROM is employed to evaluate the intermixing time, the key QoI during ladle changeover operations. RC is integrated within the POD-R ROM strategy to accurately capture the evolution of parameter-dependent temporal coefficients, particularly in scenarios with limited training data. In such cases, RC achieves competitive predictive accuracy with fewer training samples, reduced computational resources, and shorter training times compared to conventional neural network architectures. The POD-RC-ROM framework developed in this study demonstrates that the ROM predictions closely align with the full-order solution results, with slight overestimation of the intermixing time. The framework has several advantages for industrial-scale operations, offering cost-effective, time-effective, and accurate solutions for predicting complex phenomena. Furthermore, it can be effectively leveraged in design optimization processes, enabling more efficient decision-making and enhancing process control. \\

{{{The present POD-RC-ROM framework has been developed and validated for a single-strand tundish configuration, considering ladle change time as the sole varying parameter. The framework is general in nature and can, in principle, be extended to multi-strand tundishes and variable geometries. However, in such industrial scenarios, exhaustive exploration of the parameter space becomes computationally prohibitive due to the high cost of full-order simulations. In this context, reservoir computing offers an efficient data-driven framework for approximating high-dimensional non-linear dynamics. Extensions of reservoir computing, including multi-reservoir formulations and ordered aggregation strategies \cite{Li2022, YangMultiReservoir}, highlight its capability for modelling complex non-linear time-series behaviour. Future work will focus on incorporating additional operating and design parameters to enhance the robustness, generalization capability, and industrial applicability of the methodology in more complex configurations.}}

%% file: sections/ackno.tex
\section*{\large CRediT authorship contribution statement}
\textbf{Harshith Gowrachari}: Writing - original draft, Conceptualization, Data curation, Formal Analysis, Investigation, Methodology, Software, Visualization. \textbf{Mattia Giuseppe Barra}: Writing – review $\&$ editing, Resources,  Investigation. \textbf{Giovanni Stabile}: Writing – review $\&$ editing, Software, Supervision. \textbf{Gianluca Bazzaro}: Project administration, Supervision.   
\textbf{Gianluigi Rozza}: Funding acquisition, Project administration, Supervision.  

\section*{\large Declaration of Generative AI and AI-assisted technologies in the writing process}
These technologies were used to improve readability and correct spelling during the preparation of this manuscript. After using them, the authors reviewed and edited the content as needed and take full responsibility for the content of the publication. 

\section*{\large Declaration of competing interest}
The authors declare that they have no known competing financial interests or personal relationships that could have appeared to influence the work reported in this paper

\section*{\large Data statement}
Access to the data will be unavailable, as the research data includes sensitive and confidential information. 

\section*{Acknowledgements}
We acknowledge the PhD grant supported by industrial partner Danieli \& C. S.p.A. and Programma Operativo Nazionale Ricerca e Innovazione 2014-2020, P.I. Gianluigi Rozza. HG gratefully acknowledges Mattia Giuseppe Barra, Gianluca Bazzaro and Gabriele Guastaferro for their valuable discussions and for facilitating the hosting arrangements during multiple visits to Danieli Research Center (DRC).  GS acknowledges the financial support under the National Recovery and Resilience Plan (NRRP), Mission 4, Component 2, Investment 1.1, Call for tender No. 1409 published on 14.9.2022 by the Italian Ministry of University and Research (MUR), funded by the European Union – NextGenerationEU– Project Title ROMEU – CUP P2022FEZS3 - Grant Assignment Decree No. 1379 adopted on 01/09/2023 by the Italian Ministry of Ministry of University and Research (MUR) and acknowledges the financial support by the European Union (ERC, DANTE, GA-101115741). Views and opinions expressed are however those of the author(s) only and do not necessarily reflect those of the European Union or the European Research Council Executive Agency. Neither the European Union nor the granting authority can be held responsible for them. 

%% file: sections/appendix.tex
\appendix

\section{Scaling}\label{subsec:scaling}
Prior to training the RC model, the input data is normalized to the range [0,1] using Min-Max scaling, a common preprocessing step to enhance the predictive performance of neural networks. The scaling applied to the given input matrix \textit{C} is defined as:

\begin{equation}
C_{ij}^{\text{scaled}}=\frac{C_{ij}-\min\limits_{j=1, \ldots, N_s}\left(C_{ij}\right)}{\max\limits_{j=1, \ldots, N_s}\left(C_{ij}\right)-\min\limits_{j=1, \ldots, N_s}\left(C_{ij}\right)}, 
\end{equation}

\noindent where $i = 1, 2,......,N_h$, corresponds to the spatial index over the discretized domain $\Omega$, and \( {j=1,2, \ldots, N_s} \) indexes the snapshots. The computed minimum and maximum values are stored for each spatial location and subsequently used to rescale the predicted outputs back to their original physical range.

\section{Hyperparameters of RC}\label{subsec:Hyperparameter}

The key hyperparameters of the reservoir computing model used in this work are summarized in Table~\ref{table:RChyperparameters}. These values were selected through empirical tuning to minimize prediction error on the validation data.

\begin{table}[ht!]
\centering
\renewcommand{\arraystretch}{1.2} 
\begin{tabular}{|c|c|}
    \hline
    \multicolumn{2}{|c|}{\textbf{RC hyperparameters}} \\
    \hline
    Number of neurons           & 300 \\
    Activation function         & Sigmoid \\
    Leaky rate                  & 0.0023 \\
    Spectral radius             & 0.01 \\
    Regularization tolerance    & $5 \times 10^{-9}$ \\
    \hline
\end{tabular}
\caption{RC hyperparameters}
\label{table:RChyperparameters}
\end{table}